\documentclass[number,dvips]{arxstspdf}
\usepackage{graphicx}

\usepackage{flushend}
\usepackage{stfloats}

\volume{26}
\issue{4}
\pubyear{2011}
\firstpage{479}
\lastpage{501}
\doi{10.1214/08-STS263}

\makeatletter
\fnbelowfloat

\makeatother

\begin{document}
\begin{frontmatter}

\title{Analysis of the 2004 Venezuela Referendum: The Official Results
Versus the Petition Signatures}
\runtitle{Analysis of the 2004 Venezuela Referendum Results}

\begin{aug}
\author{\fnms{Gustavo} \snm{Delfino}\corref{}\ead[label=e1]{gdelfino@umich.edu}}
\and
\author{\fnms{Guillermo} \snm{Salas}\ead[label=e2]{guillermosalasd@gmail.com}}
\runauthor{G. Delfino and G. Salas}
\address{Gustavo Delfino is Professor, School of Mechanical
Engineering, Universidad Central de Venezuela, Venezuela \printead{e1}.
Guillermo Salas is Former Professor, Physics Department, Universidad
Sim\'on Bol\'\i var, Venezuela
\printead{e2}.}
\end{aug}

% ABSTRACT
%
\begin{abstract}
On August 15th, 2004, Venezuelans had the opportunity to vote in a
Presidential Recall Referendum to decide whether or not President Hugo
Ch\'avez should
be removed from office. The process was largely computerized using a
touch-screen system. In general the ballots were not manually counted. The
significance of the high linear correlation (0.99) between the number of
requesting signatures for the recall petition and the number of
opposition votes in computerized
centers is analyzed. The same-day audit was found to be not only
ineffective but a source of suspicion. Official results were compared
with the
1998 presidential election and other electoral events and distortions were
found.\looseness=-1
\end{abstract}

% KEYWORDS
%
\pdfkeywords{Referendum, election, voting machines, touch screen, ballot, correlation, uncertainty, audit}

\vspace*{-3pt}
\end{frontmatter}

%s1 ###
\section{Introduction}

A referendum to recall President Hugo Ch\'avez was carried out in
Venezuela on
August 15, 2004. The president was not recalled since the official
``no'' votes (votes
in favor of the president) exceeded the official ``s\'{\i}\hspace*{1pt}'' votes (votes in
favor of removing the
president from his post). The Organization of American States (OAS) and the
Carter Center observed the proceedings and carried out some analyses of the
voting data. They concluded that no tampering was apparent and that official
results were accurate \cite{coldaudit}.

In this manuscript, we carry out a more in-depth analysis of both the
voting data
and the data that arose from two audits carried out after the recall
referendum.\vadjust{\goodbreak}
We focus on the association between the proportion of voters who had
signed a
petition to carry out the referendum and the actual proportion of ``s\'{\i}\hspace*{1pt}'' votes
recorded at each voting center and compare what was observed relative
to what
might have been expected under some reasonable assumptions about voter
behavior. We also highlight the differences between what was observed and
what might have been expected relative to the type of voting center
(manual or
computerized) and note that official results obtained from computerized voting
centers were surprising.

We conclude that results from our analysis of the voting and auditing data
suggest that official results may not be as accurate as the OAS/Carter Center
report suggest. The objective of this article is to argue that a second
look at the
results of the Presidential Recall Referendum of 2004 in Venezuela
might be
justified.%

%s2 ###
\section{The Electoral Process in~Venezuela}

Electoral events in Venezuela are organized by the ``Consejo Nacional
Electoral''\footnote{Before the new constitution it was known as the ``Consejo
Supremo Electoral'' (CSE); see \url{http://www.cne.gov.ve}.} (CNE). On
December~6, 1998 the\vadjust{\goodbreak}
current president won the elections with 3,673,685 (57.79\%) votes versus
2,863,619 (42.21\%) votes for his adversaries. The total number of voters
in the electoral registry (REP) at that time was 11,001,913.

In 1999 a new constitution was enacted which allows citizens to request
a recall
referendum (RR) to decide whether the president should continue in office.
This referendum can only be activated after half of the period for
which the
president has been elected has transpired. In order to activate the
referendum, a
petition signed by at least 20\% of the voters registered in the REP
has to be
submitted to the CNE. It is also possible to request a consultative nonbinding
referendum with the signatures of 10\% of the voters registered in the REP.

On January 3, 2000 a new CNE was appointed but it failed to organize
elections as
scheduled. Therefore, on June 5 of 2000, yet another CNE was appointed. On
July 30, 2000 the president was reelected for a 6-year period with 3,757,773
(59.76\%) votes versus 2,530,805 (40.24\%) for his adversaries. The REP had
11,701,521 registered voters at that time.

In 2002 signatures were collected requesting a consultative referendum
which was
activated in the middle of a general national strike. The Supreme Court disabled
the CNE, therefore this consultative referendum never took place.
Citizens then
collected signatures again, this time for a recall referendum. This
was the
legal instrument which the government and the opposition represented by
the \textit{Coordinadora Democr\'atica} agreed to use, with the OAS and the Carter
Center acting as guarantors \cite{acuerdo}. This agreement ended the strike.

In 2003, the National Assembly was unable to agree on a new CNE, so the Supreme
Court appointed a new temporary CNE on August 26, 2003, even though
this procedure
was not contemplated in the constitution. The new CNE rejected the
signatures of
the petition for a referendum saying that they had been collected
before half of
the presidential period had transpired.

On November 28, 2003 signatures were collected once again, this time
under the
supervision of the CNE. On May 28, 2004, a significant fraction of the
signatures
had to be reverified by the CNE. Enough signatures were valid so, on
August 15,
2004 the Presidential Recall Referendum finally took place.

%f1 ###
\begin{figure}%[htbp]

\includegraphics{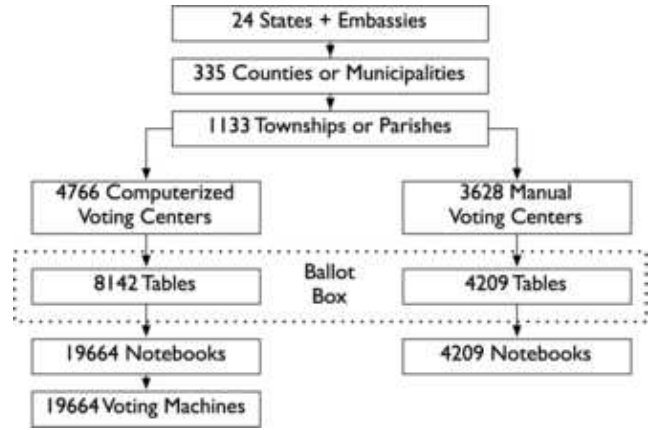}
%[scale=0.9]{layout.pdf}
\caption{Venezuelan vote collection structure.}\label{fig:layout}
\end{figure}

%f2 ###
\begin{figure*}

\includegraphics{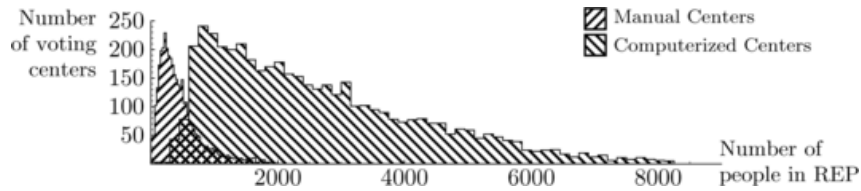}

\caption{Histogram of the size of the manual and the computerized
centers.}\label{repAutoMan}
\end{figure*}

%%%%%%%%%%%%%%%%%%%%%%%%%%%%%%%%%%%%%%%%%%%%%%%%%%%%%%%%%%%%%%%%%%%%%%%%%%%%%%%%%
%s3 ###
\section{Vote Collection Structure}

Venezuela is politically organized into states, counties
(municipalities) and
townships (parishes). Each county has one or more voting centers. There
can be
several voting tables (voting stations) per center, and each one has
one or
more electoral notebooks. In computerized centers, one voting machine is
assigned to each electoral notebook. One ballot box is assigned to each table.
Therefore, the ballots from multiple machines may be combined in a
single ballot box.
See Figure~\ref{fig:layout} for the detailed layout of the system.

Each voting center has a unique identifying code which makes it
possible to compare
electoral results on a center by center basis.

Although the number of manual centers is large, the number of people
registered in
those centers is much smaller than those registered in computerized
centers. These distributions are shown in the histograms of Figure~\ref
{repAutoMan}.

%%%%%%%%%%%%%%%%%%%%%%%%%%%%%%%%%%%%%%%%%%%%%%%%%%%%%%%%%%%%%%%%%%%%%%%%%%%%%%%%%
%s4 ###
\section{The Voting Procedure}

There were only two ways to vote\footnote{In manual voting centers it
was also
possible to cast a null vote.}: s\'{\i} (yes) or no. In order for the
president to
step down, the number of s\'{\i} votes had to be greater than 3,757,773 and
greater than the number of no votes.

Touch-screen voting machines were used for the first time in Venezuela
for the
Referendum. These machines also gave the voter a paper ballot to be
deposited in a
box. The boxes were never opened except for some of those selected for auditing.
The results were sent electronically from the voting machines to the
CNE servers
using TCP/IP connections over telephone lines, after which the voting machines
printed out the results, as well as a dupli\-cate\vadjust{\goodbreak} set of all the paper
ballots in a
continuous uncut format. The voting centers also had a
continuous
satellite TCP/IP
connection which was to be used only by fingerprint machines which were
supposed to
prevent anyone from voting twice, even in different voting centers.

In order to give the citizens confidence in the results, two audits
were made.
The first one was done on the same day as the Referendum (hot audit).
The second
one was carried out three days later (cold audit).

The official results were 3,989,008 (40,64\%) s\'{\i} votes versus 5,800,629
(59,10\%) no votes, with 14,037,900 registered voters in the REP. A
large fraction of the votes (87.1\%) were cast at computerized voting centers.

The whole electoral process and the audits were supervised and endorsed
by the OAS
and the Carter Center. They found no evidence of alterations or
tampering in the
results in their final report.

%%%%%%%%%%%%%%%%%%%%%%%%%%%%%%%%%%%%%%%%%%%%%%%%%%%%%%%%%%%%%%%%%%%%%%%%%%%%%%%%%
%s5 ###
\section{The Signatures}\label{sec:signatures}
%s5.1 ###
\subsection{Introduction}
In order to activate the Referendum, on November 28, 2003, signatures and
fingerprints were collected in a four-day event organized by the CNE, with
witnesses from all political parties. Special forms, with serial
numbers were
supplied by the CNE to all political parties. There were 2,676 signature
collection centers (SCCs), all of them in Venezuela. No signature collection
was allowed outside Venezuela.

There were two kinds of forms: types A and B. Type A forms were used in the
SCCs. Type B forms were also assigned to SCCs, but they were meant~to
be used
for house to house signature-collecting (under pro-government witness
supervision). There were 618,800 type A forms and 98,286 type B forms.
Each form had a maximum capacity of 10 signatures.\looseness=-1

The number of signatures required to activate the Referendum was 20\%
of the REP
used to elect the president, that is, $0.2\times
11{,}701{,}521=2{,}340{,}305$ signatures. The
law required the publication in a newspaper of a list of ID numbers of
all the people who signed the petition.\vadjust{\goodbreak}

The CNE divided the signatures into three categories: valid, invalid and
questionable. An important number of questionable signatures had to be
collected again in order to reach the required minimum number of signatures.

Opposition groups claimed to have submitted\break 3,467,051 signatures to the
CNE. Within the CNE, 19,842 signatures were lost.%
\footnote{See
\texttt{\href{http://buscador.eluniversal.com/2004/05/09/apo\_art\_09152D.shtml}%
{http://buscador.eluniversal.com/2004/05/09/apo\_}
\href{http://buscador.eluniversal.com/2004/05/09/apo\_art\_09152D.shtml}%
{art\_09152D.shtml}}} An additional
indeterminate number of signatures were lost before reaching the CNE.

It is reasonable to assume that most of those who signed requesting the
Referendum
intended to vote s\'{\i} in favor of the recall.\footnote{The OAS and the
Carter Certer
concur with this statement. See~\cite{CarterR1}, Section 5, second
paragraph.}
However, it is also possible that some signers voted no. This might
have been the
case for government supporters who signed the petition because they
believed they
could use the referendum to help solve the high level of political
confrontation in
the country. There were also signers who changed their political preferences
between the time of the signature collection and the vote.

In the following sections, the official results of the referendum will
be compared
with the signatures collected. This will reveal some important facts
about these
results.

% Machado asegur que cuentan con 3 millones 467 mil 51 firmas para
%solicitar el
% revocatorio. Insisti en dejar claro que se obtuvo ms de un milln
%(1.011.050)
% de firmas por encima del requisito constitucional, que exige 2,4
%millones.

% Oficialmente se perdieron 19.842
% http://buscador.eluniversal.com/2004/05/09/apo_art_09152D.shtml

% 3 467 051 - 19 842 = 3 447 209

% firmas en mi base de datos = 3 448 747

%%%%%%%%%%%%%%%%%%%%%%%%%%%%%%%%%%%%%%%%%%%%%%%%%%%%%%%%%%%%%%%%%%%%%%%%%%%%%%%%%
%s5.2 ###
\subsection{Si Vote Uncertainty With Regard to Signatures}\label
{sec:uncertainty}

Let $k$ be the relative number of s\'{\i} votes, as defined in
equation~(\ref{eq:k}):
%
%e1 ###
\begin{equation}
k=\frac{\mbox{s\'{\i} votes}}{\mbox{signatures}}.
\label{eq:k}
\end{equation}

%f3 ###
\begin{figure*}[t]

\includegraphics{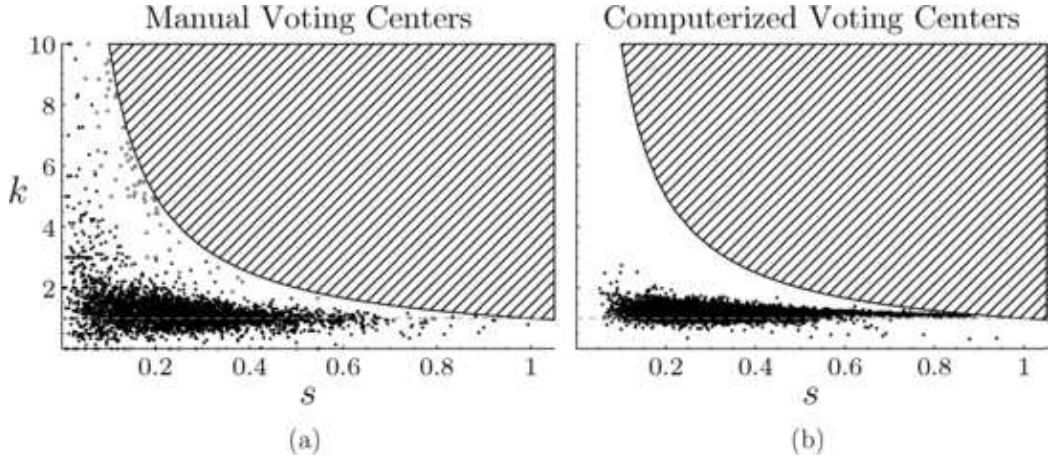}

\caption{Relationship between $k$ and $s$ for computerized and manual centers.
The shadowed area contains the mathematically impossible values of $k$.
The maximum $k$ value is $1/s$. The hollow dots represent voting centers
located in consular offices.}\label{fig:laurent}
\vspace*{3pt}
\end{figure*}

Also, let $s$ be the relative number of signatures in a voting center,
as defined in equation~(\ref{eq:s}):
%
%e2 ###
\begin{eqnarray}\label{eq:s}
s&=&\frac{\mbox{signatures}}{\mbox{s\'{\i} votes $+$ no votes $+$ null
votes}}\nonumber\\[-8pt]\\[-8pt]
&=&\frac{\mbox{signatures}}{\mbox{total votes}}.\nonumber\vadjust{\goodbreak}
\end{eqnarray}

For each value of $s$, there is a maximum possible~$k$ which is just
$1/s$ as shown in equation~(\ref{eq:kmax}):
%
%e3 ###
\begin{eqnarray}\label{eq:kmax}
\ensuremath{k_{\mbox{\tiny max}}}&=&
\frac{\max(\mbox{\small s\'{\i} votes})}{\mbox{signatures}}=
\frac{\mbox{total votes}}{\mbox{signatures}}\nonumber\\[-8pt]\\[-8pt]
&=&
\frac{\mbox{total votes}}{s\cdot\mbox{total votes}}=\frac{1}{s}.\nonumber
\end{eqnarray}

In voting centers with a large value of $s$, we expected a value of $k$
around 1. This is because each signature has a high probability of
resulting in
a s\'{\i} vote, and at the same time \ensuremath{k_{\mbox{\tiny max}}}\
gets close to 1.

For example, in a voting center with $1{,}000$ total votes and 900
signatures, the
number of expected s\'{\i} votes is between 900 and $1{,}000$. Here
$s=900/1{,}000=0.9$ and
$\ensuremath{k_{\mbox{\tiny max}}}=1/0.9=1.1\bar{1}$. Therefore, the
uncertainty in the value of $k$ is
very small, as it should be between\footnote{The value of $k$ could be
lower than 1 if, for any reason, the number of votes was low (e.g.,
high abstention).} 1~and $1.1\bar{1}$.

The situation is completely different in voting centers with a small
value of
$s$. Notice that there is an essential singularity in $k$ at $s=0$ as
shown in
equation~(\ref{eq:singularidad}):
%
%e4 ###
\begin{equation}
k=\frac{ \mbox{s\'{\i} votes} / \mbox{total votes} }{s}.
\label{eq:singularidad}
\end{equation}
This singularity can produce very high values of $k$ in the
neighborhood of $s=0$. Hence, the level of uncertainty in $k$ becomes
very large.

For example, in a voting center with $1{,}000$ total votes and 2
signatures, the
number of expected s\'{\i} votes is between 2 and $1{,}000$. Here
$s=2/1{,}000=0.002$ and
$\ensuremath{k_{\mbox{\tiny max}}}=1/0.002=500$. Therefore, the
uncertainty in the value of $k$ is
extremely large, as it should be between 1 and 500.

%f4 ###
\begin{figure*}%[htbp]

\includegraphics{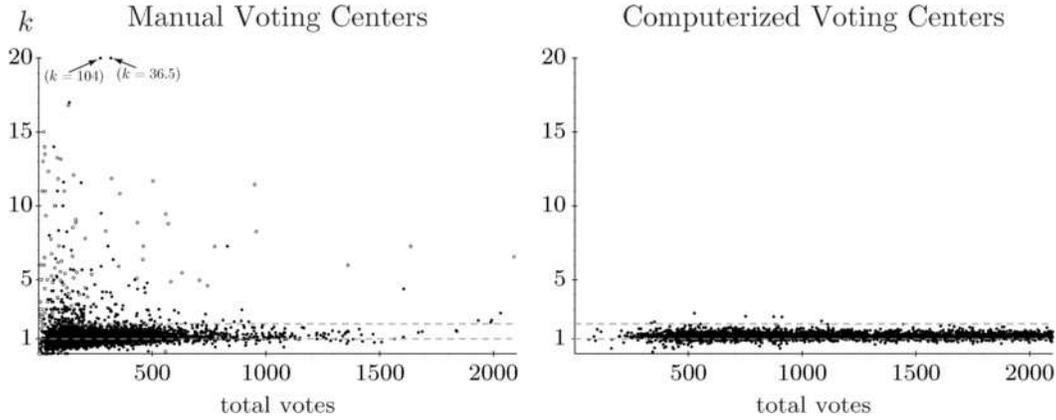}%
%[scale=1]{k_vs_size.pdf}
\caption{Relationship between $k$ and total number of votes for computerized
and manual centers in the same size range. The hollow dots represent voting
centers located in consular offices.}\label{fig:k_vs_size}
\end{figure*}

%f5 ###
\begin{figure*}[b]
\includegraphics{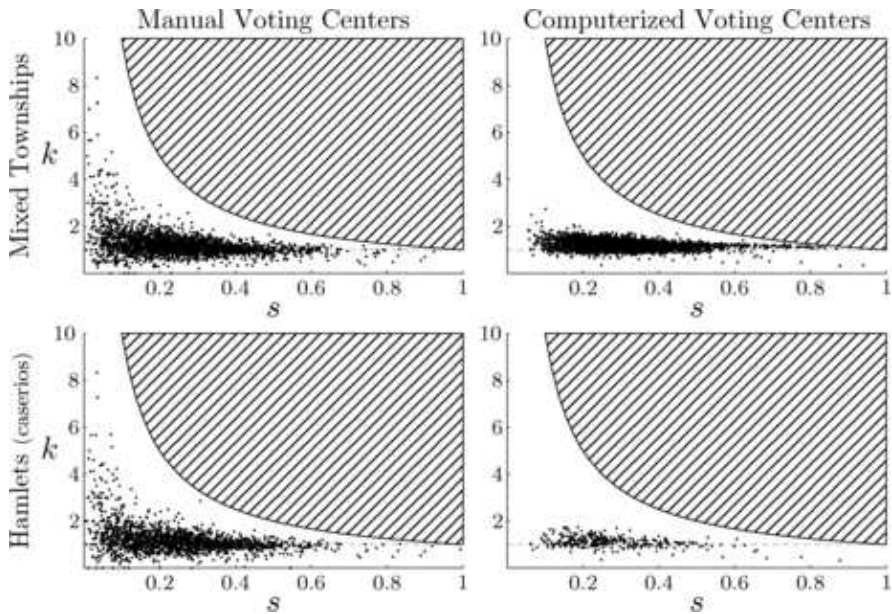}%
%[scale=1]{caserios.pdf}
\caption{Relationship between $k$ and $s$ for computerized (right) and manual
centers (left) for mixed townships (top) and hamlets
(bottom).}\label{fig:caserios}
\end{figure*}

The reasons for the uncertainty in $k$ just discussed are purely
mathematical. In
practical terms, high values of $k$ in centers with a small $s$ were
due to
the following facts:

\begin{itemize}

\item There were only $2{,}676$ SCCs compared to $8{,}394$ voting centers.
Therefore,
voters living far from a SCC could not sign the petition, even if they wanted
to. This was the case in mostly rural areas.

\item There were many people who did not sign the petition because of
their fear
of retribution from the government. On the other hand, voting was
secret.%

\item There were s\'{\i} votes from people who could not sign because they were
not in the REP or were outside the country at the time of signature
collection.

\item Some SCCs ran out of forms. Not everyone was able to go to a more
distant SCC to sign.

\item An undetermined number of signatures were lost.

\item There were s\'{\i} votes from people who just didn't bother to sign the
petition.

\end{itemize}

Notice that all these issues with the signatures did not affect all voting
centers equally. Centers with a~small value of $s$ are more likely to have
been affected by these issues than centers with a high value of~$s$.

A plot of $k$ versus $s$ is shown in Figure~\ref{fig:laurent}. Notice
that when
$s$ is not large, \textit{all} the computerized centers are very far away from
$k_{\mbox{\tiny max}}$, clearly contradicting the expected nonlinear behavior
with respect to~$s$. On the other hand, the manual center results are
effectively distributed in the allowed range regardless of the relative number
of signatures.

In summary:
\vspace{8pt}

\framebox{
\begin{minipage}[t]{200pt}{
It is expected that $k$'s from voting centers with a small value of $s$ will
be much more variable than those with large values of $s$.}
\end{minipage}
}

\vspace{8pt}

%s5.2.1 ###
\subsubsection{\texorpdfstring{Behavior of $k$ with regard to the size and~cha\-racteristics of
the voting centers.}{Behavior of k with regard to the size and characteristics of the voting centers}}

Although the manual centers tend to have fewer voters than the computerized
centers, this does not seem to be the only reason for the different behavior
in $k$. This can be seen in Figure~\ref{fig:k_vs_size}.

There were many small computerized voting centers in rural areas. Many
used mobile
phone lines to connect the voting machines to the CNE servers to
transmit the
results because of the lack of regular phone lines in these remote areas.

%f6 ###
\begin{figure*}

\includegraphics{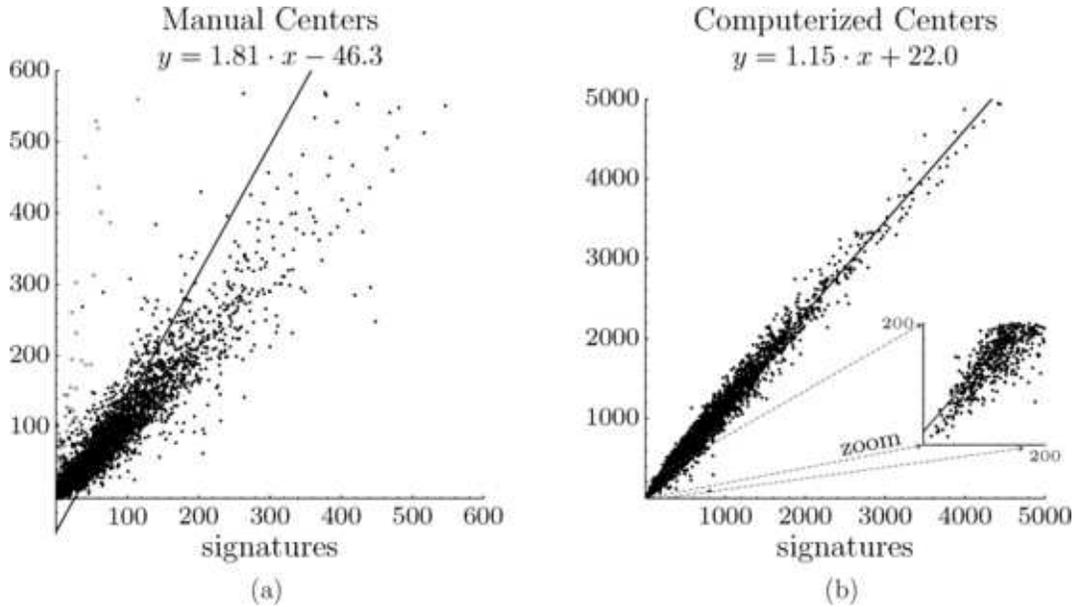}

\caption{Manual centers have a correlation of 0.607 with respect to the
signatures while computerized centers have a correlation of 0.989. A
correlation of 1 would look like a straight
line.}\label{fig:votesVsSignatures}
\vspace*{-3pt}
\end{figure*}

There were 586 townships which included both manual and computerized
voting centers. These\break mixed townships had 5,449 voting centers
(2,538 manuals
and 2,911 computerized). Notice in Figure~\ref{fig:caserios} (top) that
the behavior of $k$ in these mixed townships, is very different for
manual and
computerized centers. Appendix~\ref{apndx:google} shows an example of
such a mixed township.

%In Figure~\ref{fig:caserios} (top) is observed that
%the behavior of $k$ in these mixed townships, is very different for
%manual and
%computerized centers.

Another interesting comparison is related to hamlets (``caser\'\i
os''). A total of 2,162 voting centers in hamlets were
identified\footnote{The
official list of voting centers was searched for the word ``CASERIO''
in the
address field. These produced the list of 2,162 voting centers.}
(1,852 manual
and 310 computerized).

Due to the reasons mentioned in Section~\ref{sec:uncertainty}, many hamlets
must have been far away from a SCC. For this reason voting centers
located in
hamlets should include large values of $k$. In Figure~\ref{fig:caserios}
(bottom) it can be seen that these large values are found only in manual
voting centers.

Furthermore, Figure~\ref{fig:caserios} shows that the behavior of the
$k$ values in
computerized voting centers in hamlets looks more like that of the rest
of the
computerized centers than the behavior of the 1,852 manual centers
located in the
rest of the hamlets.

%%%%%%%%%%%%%%%%%%%%%%%%%%%%%%%%%%%%%%%%%%%%%%%%%%%%%%%%%%%%%%%%%%%%%%%%%%%%%%%%%
%s5.3 ###
\subsection{Correlations Between Si Votes and Requesting Signatures}

%The correlation $r$ is a measure of {\em linear} association between
%two variables. If the linear association is very strong then $r$ is
%close to 1. If it is weak, then $r$ is close to 0. A value of -1
%indicates a mirror dependance which wouldn't make sense in our context.
%Correlations are calculated by using Pearson's method as shown in
%Equation~\ref{eq:correlation}.
%r=\sum_{i}{\frac{(x_i-\bar{x})(y_i-\bar{y})}{ \sqrt{\sum_{i}{(x-

Let \ensuremath{r_{\!\mbox{\tiny s\'{\i}}}}\ be the correlation of s\'{\i}
votes with respect to the number of
signatures.

The Carter Center and the OAS said the following in one of their
reports \cite{CarterR1}:

\begin{quote}
A very high correlation between the number of signers and the number of
s\'{\i}\vadjust{\goodbreak}
votes per center in the universe of automated voting machines has been
found---a correlation coefficient of 0.988. This means that in voting centers
where a high signer turnout was obtained, a high s\'{\i} vote also was
obtained.\footnote{This correlation value was reproduced with a
difference of
just 0.001 which is negligible.}
\vspace*{-2pt}
\end{quote}

What this report does not mention is that for manual voting centers, the
correlation is 0.607,\break a~much lower value. This difference can be
visualized in
Figure~\ref{fig:votesVsSignatures}. Notice that a straight line from the
origin to each of the points has a slope of $k$. The high correlation value
for computerized centers translates into similar $k$ values (or slopes) for
most centers.

%f7 ###
\begin{figure*}[b]

\includegraphics{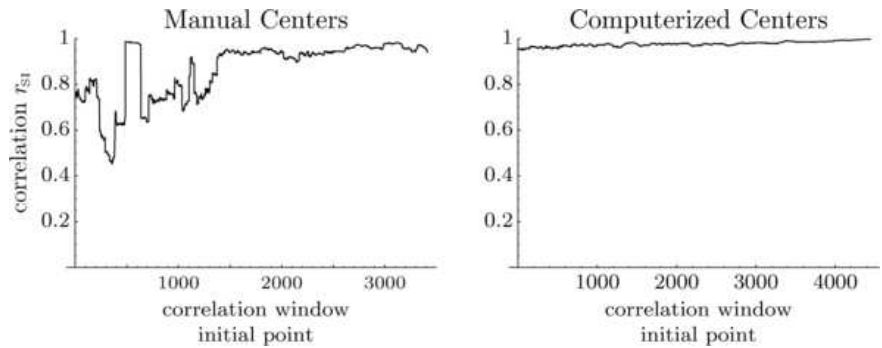}

\caption{Correlations plot using a window of 150 voting
centers.}\label{fig:ama_lin}
\end{figure*}

In this case, the high correlation in computerized voting centers also implies
that in voting centers where a low signer turnout was obtained, a low
s\'{\i} vote
was also obtained. This can be seen in the origin of
Figure~\ref{fig:votesVsSignatures}(b). Hence, when the number of signatures
tends to zero, the number of s\'{\i} votes also tends to zero. But, as
observed in
Figure~\ref{fig:votesVsSignatures}(a), manual centers do not exhibit the same
behavior.

The behavior found in computerized centers seems unexpected because the
relationship between signatures and s\'{\i} votes should not be linear,
especially when
the number of signatures is small. As explained in\vadjust{\goodbreak} Section~\ref
{sec:uncertainty},
you could expect a large number of s\'{\i} votes if there were a large
number of
signatures, but as the number of signatures per center decreases, the
level of
uncertainty in the number of s\'{\i} votes with respect to the number of signatures
increases.\looseness=1

In Table~\ref{table:rsi} the correlations are calculated for centers where
signers were a minority ($s\le0.5$) and a majority ($s>0.5$). Notice
that as
expected, the correlation for manual centers is much higher when there are
many signatures (0.947) than when there are fewer signatures (0.613).
This is
the expected behavior because when you have many signatures the
uncertainty of
$k$ is small, and the number of s\'{\i} votes is equal to
$k\times\mbox{signatures}$ so the uncertainty in the absolute number
of s\'{\i}
votes is also small.\looseness=1

%t1 ###
\begin{table}%[htdp]
\tabcolsep=0pt
\caption{Correlations of s\'{\i} votes with respect to the relative
number~of~signatures $s$ per center, for manual and computerized voting
centers}\label{table:rsi}
\begin{tabular*}{240pt}{@{\extracolsep{\fill}}lcccccc@{}}
\hline
& \multicolumn{2}{c}{$\bolds{\mathit{s \le\mathrm{0.5}}}$}& \multicolumn{2}{c}{$\bolds{\mathit{s} > \mathrm{0.5}}$}& \multicolumn{2}{c}{\textbf{All}}
\\[-6pt]
& \multicolumn{2}{c}{\hrulefill}& \multicolumn{2}{c}{\hrulefill}&\multicolumn{2}{c@{}}{\hrulefill}\\
& \textbf{\ensuremath{r_{\!\mbox{\tiny s\'{\i}}}}}& \multicolumn{1}{c}{\textbf{\#}} &\textbf{\ensuremath{r_{\!\mbox{\tiny s\'{\i}}}}}& \textbf{\#}
& \textbf{\ensuremath{r_{\!\mbox{\tiny s\'{\i}}}}}& \textbf{\#} \\
\hline
Manual & 0.613 & 3,375 & \textbf{0.947} & 221 & 0.607 & 3,596 \\
Computerized & 0.983 & 3,943 & 0.994 & 645 & 0.989 & 4,588 \\
Both & 0.953 & 7,318 & 0.996 & 866 & 0.973 & 8,184 \\
\hline
\end{tabular*}
\end{table}

In the case of the 645 computerized voting centers where \mbox
{$s>0.5$} the
correlation was 0.994 which is very high. It stands out that in the
computerized voting centers where signers were a minority, the
correlation is
still very high at 0.983. Furthermore, there is not a single computerized
voting center with many more s\'{\i} votes than signatures as seen in
Figure~\ref{fig:votesVsSignatures}(b). In other words, for some reason,
computerized centers do not seem to show the expected nonlinear relationship
between signatures and s\'{\i} votes.

%%%%%%%%%%%%%%%%%%%%%%%%%%%%%%%%%%%%%%%%%%%%%%%%%%%%%%%%%%%%%%%%%%%%%%%%%%%%%%%
%s5.4 ###
\subsection{Correlation Plot}

In order to further investigate the change of uncertainty as the
relative number of signatures varies, a~technique similar to a moving
average is used. The~difference is that instead of calculating an
average, a correlation is calculated. A window size of 150 voting
centers was used. This is the same number of centers that were
audited.

%f8 ###
\begin{figure*}[t]

\includegraphics{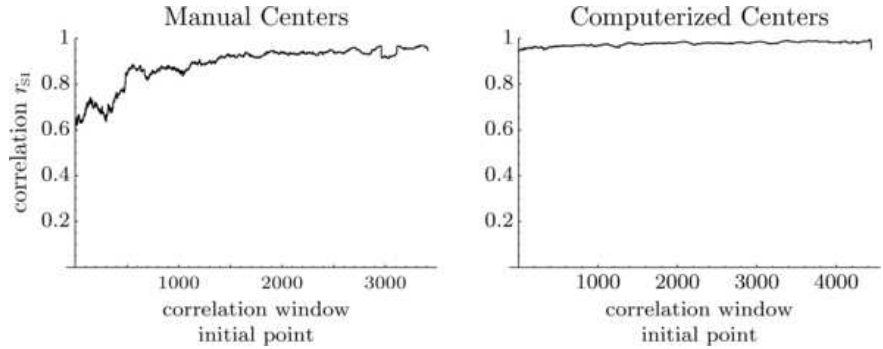}
%[]{ama_log.pdf}
\caption{Correlations plot (logarithmic scale) using a 150 voting centers
window.}\label{fig:ama_log}
\end{figure*}

In order to do this, the first step is to sort the voting centers,
computerized and manual, according to their $s$ value. Then \ensuremath
{r_{\!\mbox{\tiny s\'{\i}}}}\ is
calculated for centers in positions 1 to 150. Subsequently \ensuremath
{r_{\!\mbox{\tiny s\'{\i}}}}\ is calculated
for centers in positions 2 to 151, and so on. The result is shown in
Figure~\ref{fig:ama_lin}.

For manual centers, there are large variations in the correlation in
the left side of Figure~\ref{fig:ama_lin}. This is the result of
outliers coming in and out of the 150 centers calculation window. As
the outliers are real official data, they should not be
dropped. Instead, logarithms can be used for both the number of votes
and signatures. This way the effect of the outlier is taken into
account in a better way. The result of using this technique is shown
in Figure~\ref{fig:ama_log}.\looseness=1

Regardless of whether correlations are calculated on a linear scale
(Figure~\ref{fig:ama_lin}) or on a logarithmic scale
(Figure~\ref{fig:ama_log}), the important fact to point out is that the
reduction in correlation as $s$ decreases is large for manual centers, whereas
it is negligible for computerized centers.

%s6 ###
\section{The Hypothesis}

What has been presented thus far should be enough to cast a serious
shadow of
doubt regarding the official results in the computerized centers. Based
on this, it is natural to consider the following hypothesis\footnote{The
mechanics of how votes could have been altered, and by whom is not studied
here. However, the fact that the machines established a TCP/IP
connection to
the CNE, disconnected and only then printed the results, opens many security
holes. These issues are beyond the scope of this article.}:

\begin{hypothesis*}
In computerized centers, official\break \mbox{results} were forced to follow a linear
relationship with respect to the number of signatures.
\end{hypothesis*}

If this hypothesis were true, because of the reasons explained in
Section~\ref{sec:uncertainty}, the results would be distorted with
respect to
reality, especially in voting centers with a small $s$ value.

In places where the signatures did not correctly capture the political
intention of the people, two things would happen:

\begin{enumerate}
\item The number of s\'{\i} votes, according to the official CNE results, would
tend to be much lower than the number of real s\'{\i} votes.
\item The official results of those computerized voting centers would
be a
poor representation of the political intentions in the area.
\end{enumerate}

In the next section the results of the referendum will be compared to
those of
the 1998 presidential election in order to find out if these
distortions are
indeed present.

% _ ___ ___ ___
% / |/ _ \ / _ \ ( _ )
% | | (_) | (_) |/ _ \
% | |\__, |\__, | (_) |
% |_| /_/  /_/    \___/

%s7 ###
\vspace*{3pt}
\section{1998 Election Comparison}\label{sec:1998}
\vspace*{3pt}

Despite the fact that more than 5 years separate the 1998 presidential
election and the Referendum, and that the Referendum was not an election,
there are reasons that make the comparison of both events interesting:

\begin{itemize}
\item In both cases the future of the presidency was at stake.
\item In Venezuela, since 1958 a new president had been elected every 5 years.
Immediate reelection was prohibited by the 1961 constitution. Between
the 1998
election and the 2004 Referendum, 5 years and 8 months had gone by. On
the other
hand, the president had repeatedly claimed that he would stay in office at
least until the year 2021.
\item Both events were open for all Venezuelan citizens in the electoral
registry.
\item Both cases involved a very polarized electorate. In 1998 the top two
candidates obtained 96.17\% of the valid votes. The other 3.83\% of the votes
went to candidates who were also politically opposed to the winning candidate.
\item There were 8,431 voting centers in 1998 and 8,394 voting
centers for the
Referendum. The events had 8,328 voting centers in common.
\item Comparing the 1998 election and the Referendum results gives an estimate
of whether the popularity of the president increased or decreased in
the vicinity
of each voting center.
\end{itemize}

Additionally, the 1998 electoral results are used for comparison
because at
that time, the CNE was not under the influence of the current government.

%f9 ###
\begin{figure*}%[htbp]

\includegraphics{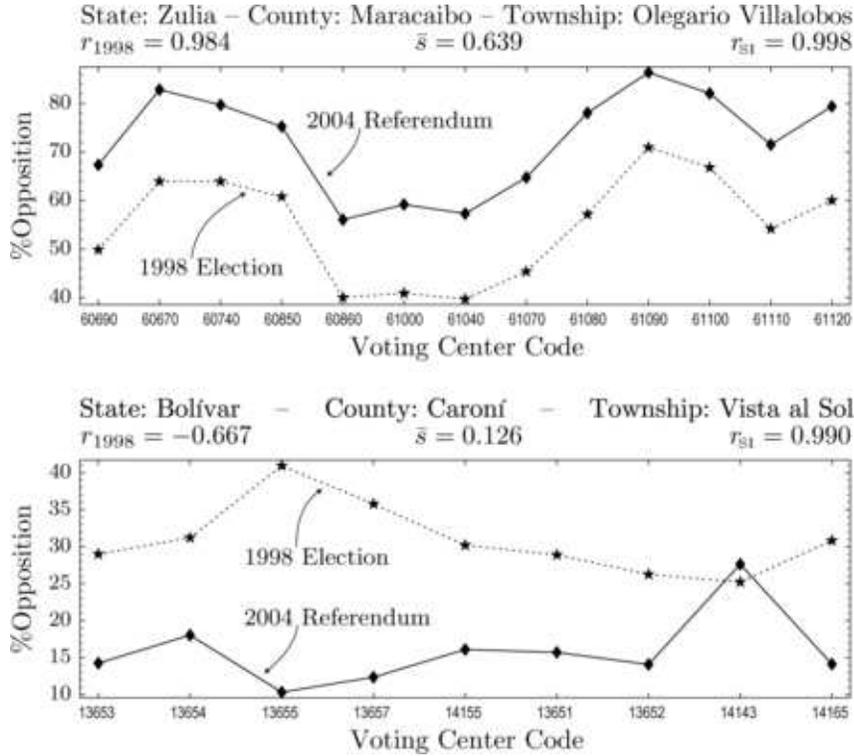}
%{parroquia.pdf}
\caption{Two sample townships. All the centers shown are computerized voting
centers.}\label{fig:parroquia}
\vspace*{3pt}
\end{figure*}

%%%%%%%%%%%%%%%%%%%%%%%%%%%%%%%%%%%%%%%%%%%%%%%%%%%%%%%%%%%%%%%%%%%%%%%%%%%%%%%%%
%s7.1 ###
\subsection{Correlations Between \% of Opposition Votes in 1998 and in RR}

By comparing the electoral results (percentage of opposition) on a
township by
township basis, it was detected that some of them had a high
correlation with
respect to previous results while others had a~very low correlation.
The townships with higher opposition results with
respect to 1998 tend to have a~higher correlation than the others.
This correlation will be called $r_{1998}$, and the\vadjust{\goodbreak}
percentage of opposition difference will be called \ensuremath{\Delta
\%_{1998}^{\mbox{\scriptsize RR}}}\ as defined in
equation~(\ref{delta98RR}):
%
%e6 ###
%e5 ###
\begin{eqnarray}\label{delta98RR}
\ensuremath{\Delta\%_{1998}^{\mbox{\scriptsize RR}}}&=& (\%\ \mbox
{Opposition in RR})\nonumber\\[-8pt]\\[-8pt]
&&{}-(\%\ \mbox{Opposition in 1998}).\nonumber
\end{eqnarray}

In order to illustrate this, the results of two townships are plotted in
Figure~\ref{fig:parroquia}. In the \textit{``Olegario Villalobos''}
township, the
correlation with respect to the signatures and the 1998 percentage of
opposition is large at \mbox{$\ensuremath{r_{\!\mbox{\tiny
s\'{\i}}}}=0.988$} and $r_{\scriptsize1998}=0.984$
respectively. Additionally, notice that the average $s$ is 0.639, so signers
were the majority in this township. Therefore, the signatures are
likely to
have captured the political intentions of voters here.

In the case of the \textit{``Vista al Sol''} township, the average
$s$ is very
low. Therefore, the uncertainty in the number of s\'{\i} votes with respect
to the
signatures could be large, as was shown in Section~\ref
{sec:uncertainty}. In
other words, the signatures are not likely to have captured the political
intentions of the township accurately. This uncertainty is just not
seen in
the official results, as the correlation of s\'{\i} votes with respect to the
signatures is~0.990. Furthermore, the referendum results seem very distorted
with respect to the\vadjust{\goodbreak} 1998 election, with a negative correlation of
$-$0.667. In
this township, the center with the most opposition in 1998 ended up
being the
most pro-government, and vice versa.\footnote{This center returned to being
the one with the most opposition 77 days later in the state governors
election.}

The two townships shown in Figure~\ref{fig:parroquia} behave
consistently with
the hypothesis. \textit{``Olegario Villalobos''} was able to increase its
percentage of opposition because many signatures were collected,
whereas \textit{``Vis\-ta al Sol''} could not increase its percentage of opposition
because only
a few signatures were collected. If this repeats itself in the rest of the
country, then $r_{\scriptsize1998}$ would be large when \ensuremath
{\Delta\%_{1998}^{\mbox{\scriptsize RR}}}\ is large,
and $r_{\scriptsize1998}$ would be small when \ensuremath{\Delta\%
_{1998}^{\mbox{\scriptsize RR}}}\ is small. In an
untouched process, these two variables should be independent.

%f10 ###
\begin{figure*}[t]

\includegraphics{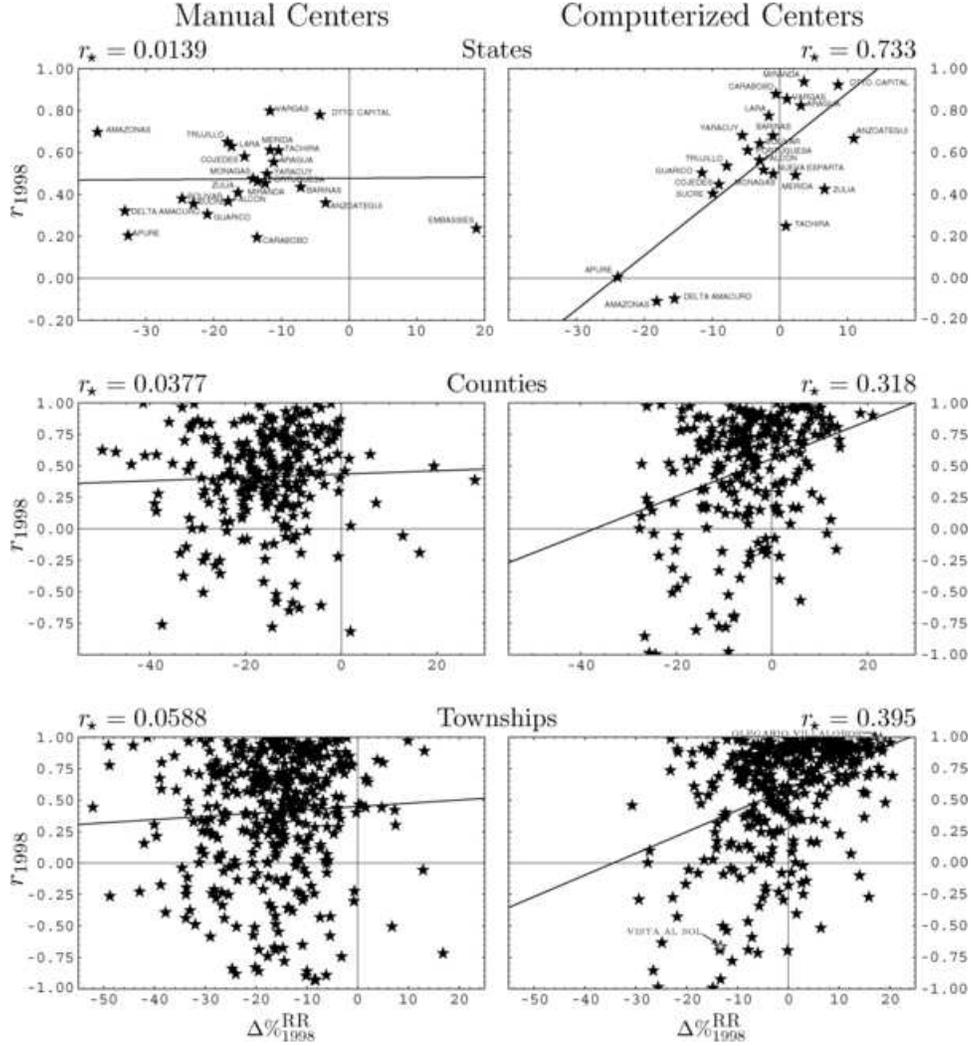}
\vspace*{-3pt}
\caption{Relationship between $r_{1998}$ and \ensuremath{\Delta\%
_{1998}^{\mbox{\textup{\scriptsize RR}}}}\ at the state, county and
township levels for manual and computerized voting centers. The correlation
between \ensuremath{\Delta\%_{1998}^{\mbox{\textup{\scriptsize RR}}}}\ and
$r_{1998}$ is shown as $r_{\!\star}$. Assuming that
\ensuremath{\Delta\%_{1998}^{\mbox{\textup{\scriptsize RR}}}}\ and $r_{1998}$
are independent, the probability of seeing those
$r_{\!\star}$ values is calculated in
Appendix~\textup{\protect\ref{apndx:mcarlo}}.}\label{fig:six}\vspace*{-4pt}
\end{figure*}

In Figure~\ref{fig:six}, it is shown that, indeed in all of the
country there
is a strong relationship between \ensuremath{\Delta\%_{1998}^{\mbox
{\scriptsize RR}}}\ and $r_{\scriptsize1998}$ for
computerized centers at the township, county and state levels. This
relationship is much weaker---almost inexistant---for manual voting centers.
This finding is consistent with the hypothesis.

%s8 ###
\section{Variability in Values of $\lowercase{k}$ and the Correlation Between Percentage
of Opposition and Values of $\lowercase{s}$, for Various Electoral
Events}\vspace*{-3pt}\label{sec:rDiff}

In Section~\ref{sec:uncertainty}, it was stated that as the value of $s$
decreases, the variability in $k$ is expected to increase. According to
equation~(\ref{eq:singularidad}) this variability must also be present
in the
relation between $s$ and the percentage of opposition. Therefore, as $s$
becomes small, it should correlate poorly with the percentage of opposition.
For this reason, when $s$ is small, it should not determine the
percentage of
opposition. On the other hand, when $s$ becomes large, it should correlate
better with the percentage of opposition.

Let $r_{\!s}$ be the correlation of the percentage of opposition and
$s$, and\vadjust{\goodbreak}
let $\tilde{s}$ be the median of all the values of $s$ for computerized
centers. For the subset of computerized centers with $s\le\tilde{s}$ this
correlation will be called $r_{\!s,s\le\tilde{s}}$, and for the remaining
centers where $s>\tilde{s}$ the correlation will be called
$r_{\!s,s>\tilde{s}}$.

The value of $r_{\!s,s\le\tilde{s}}$ should be smaller than
$r_{\!s,s>\tilde{s}}$. These properties just defined are calculated for
various electoral events in Table~\ref{table:rDiff}.

%t2 ###
\begin{table*}[t!]
\caption{Correlation $r_{\!s}$ for computerized centers with $s$ above and
below $\tilde{s}$, for different electoral events}\label{table:rDiff}
\begin{tabular*}{\textwidth}{@{\extracolsep{\fill}}lcccc@{}}
\hline
\textbf{Date} & \textbf{Event} & $\bolds{r_{\!s,s\le\tilde{s}}}$ & $\bolds{r_{\!s,s>\tilde{s}}}$ &
$\bolds{r_{\!s,s>\tilde{s}}-r_{\!s,s\le\tilde{s}}}$ \\
\hline
Dec 6, 1998 & Presidential election & 0.439 & 0.685 & \phantom{$-$}0.246 \\
 Jul 30, 2000 & Presidential election & 0.607 & 0.802 & \phantom{$-$}0.195 \\
 Aug 15, 2004 & Referendum official results & \textbf{0.845} &
\textbf{0.830} & \textbf{$-$0.015} \\
 Aug 15, 2004 & Exit polls & 0.325 & 0.739 & \phantom{$-$}0.414 \\
 Oct 31, 2004 & States Governors election & 0.475 & 0.707 & \phantom{$-$}0.232\\
 \hline
\end{tabular*}
\end{table*}

The exit poll shown in Table~\ref{table:rDiff} was made under the
supervision of Penn, Schoen and Berland Associates.

The State Governors election took place just\break 77~days after the
Referendum. By
counting votes for and against the pro-government candidate, a
percentage of
opposition was calculated. During this election, the same voting
machines were
used, but there was an important difference: the paper ballots were manually
counted for a randomly selected voting machine in each and every voting
center. The results for the correlation $r_{\!s}$ for this election are shown
in Table~\ref{table:rDiff}.

\begin{figure*}

\includegraphics{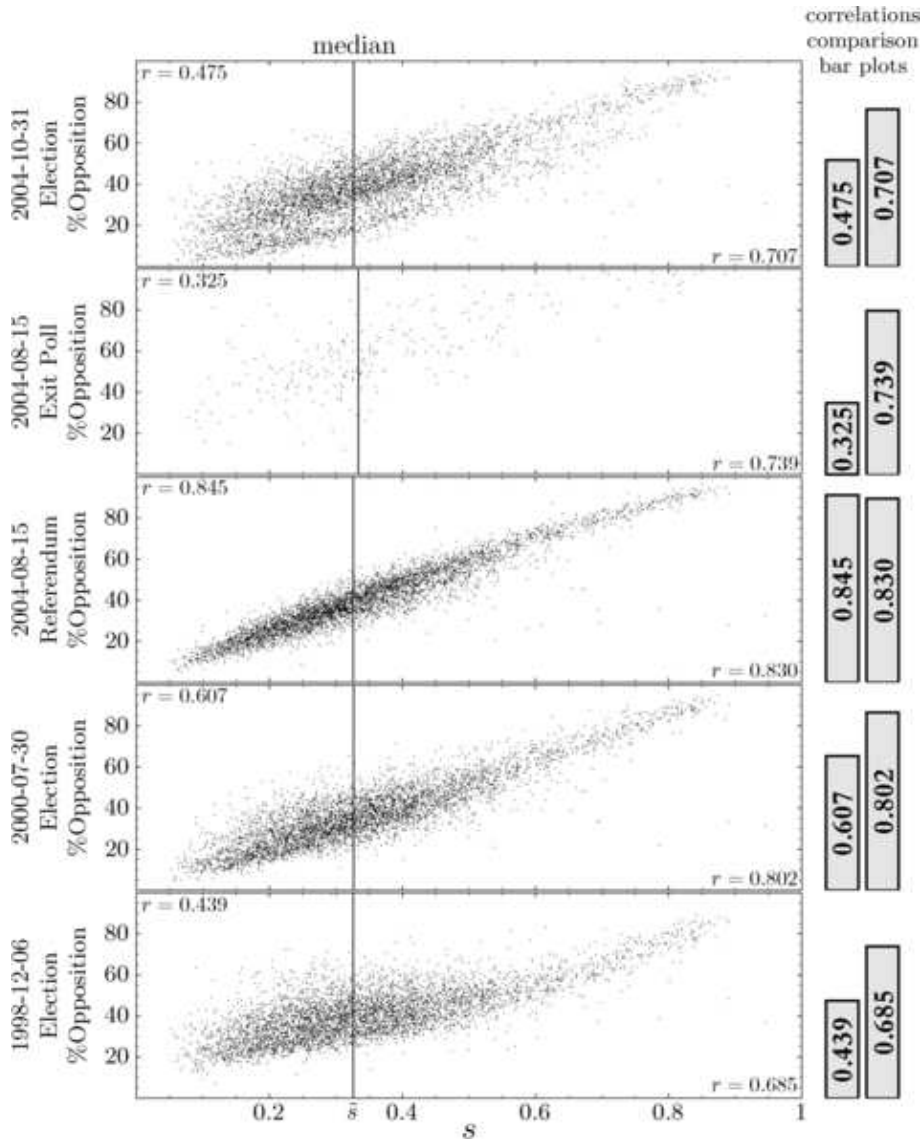}%
\vspace*{-4pt}%[scale=1]{gUncertainty.pdf}
\caption{Correlation between percentage of opposition and $s$ for the lower
two ($s\le\tilde{s}$) and upper two quartiles ($s>\tilde{s}$) for
computerized
centers. The correlation for the lower two quartiles is expected to be smaller
than the correlation in the upper two quartiles. This expected
difference is
not seen in the Referendum official results.}\label{fig:uncertainty}\vspace*{-4pt}
\end{figure*}

From Table~\ref{table:rDiff} it is clear that only the Referendum official
results fail to exhibit a positive correlation difference. Also notice in
Figure~\ref{fig:uncertainty} that for the Referendum official results,
there is
not a single voting center with a small $s$ and large percentage of
opposition. The fact that only in the official Referendum results
$r_{\!s,s\le\tilde{s}}$ is not smaller than $r_{\!s,s>\tilde{s}}$ is
consistent with the hypothesis.

%El hecho de que la fraccin de firmantes ($s$), de la mitad de los
%centros electorales
%computarizados con valores ms bajos de $s$, hallan determinado en el
%referndum a la
%fraccin opositora en un grado igual e incluso mayor que el resto de
%los centros
%electorales computarizados, es, por decir lo menos, desconcertante.

% _ _ _ _ _ _ _
% | | | | ___ | |_   / \ _ _ __| (_) |_
% | |_| |/ _ \| __| / _ \| | | |/ _` | | __|
% |     | (_) | |_ / ___ \ |_| | (_| | | |_
% |_| |_|\___/ \__| /_/ \_\__,_|\__,_|_|\__|
%s9 ###
\section{Hot Audit}

In general, the paper ballots from the computerized centers were not manually
counted. The CNE assured the Venezuelan citizens that the voting
machines had
to accurately reflect the voters intention, because a sample of 192 machines
(1\% of them) would be randomly selected and audited the same day of the
referendum. This is indeed a valid way of eliminating suspicion, as
long as
the selection is a truly random sample of \textit{all} the voting machines.

%f11 ###

%f12 ###
\begin{figure*}[t]

\includegraphics{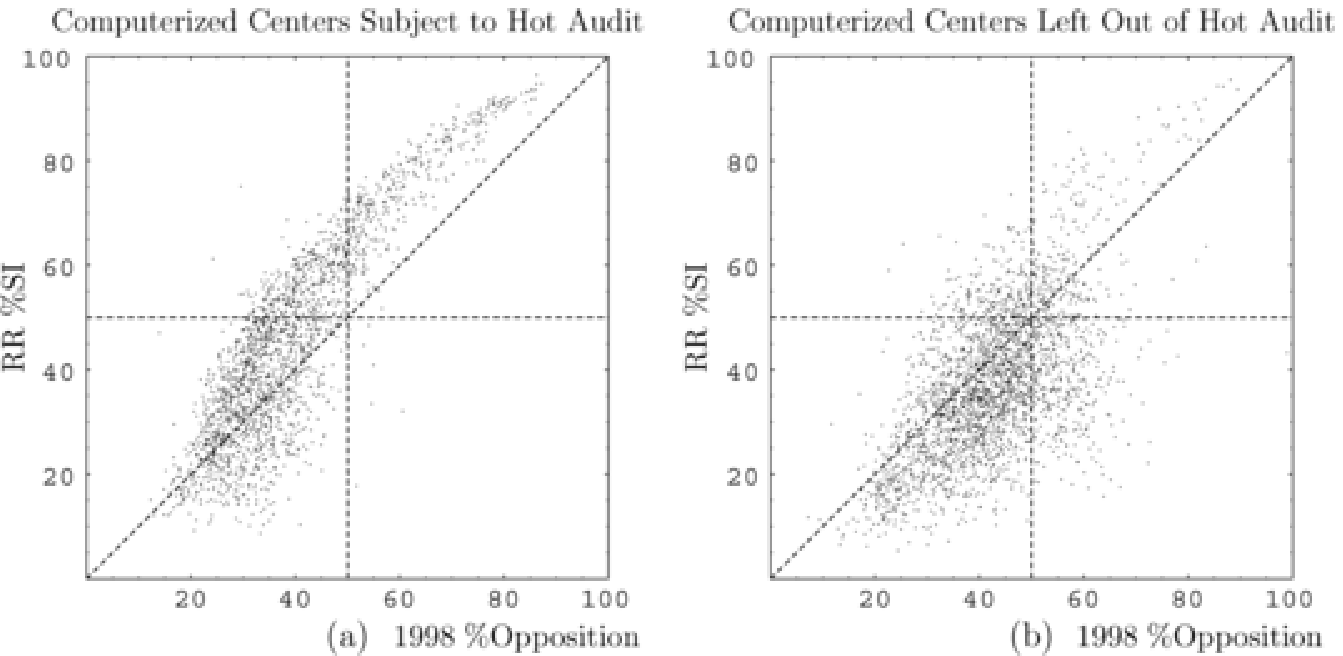}\vspace*{-1pt}
%[scale=0.95]{phaseHotAudit.pdf}
\caption{Centers inside \textup{(a)} and outside \textup{(b)} of the 20 counties to
where the
hot-audit drawing was restricted. }\label{fig:20counties}
\vspace*{-3pt}
\end{figure*}

The day of the referendum, the CNE informed the public that because of
logistical reasons, the sample would be taken from a restricted
universe of 20
counties located in urban areas, leaving out of the audit more than 300
counties. With this decision, confidence in the results was adversely
affected to
say the least.

The computerized voting centers inside and outside of the 20 counties,
to which
the hot-audit universe was reduced, are shown in Figure~\ref{fig:20counties}.
It is clear that these 20 counties are not representative of all the
computerized voting centers. See Appendix~\ref{apndx:k20counties} for further
details on this subject.

%f13 ###
\begin{figure*}[b]
\vspace*{-3pt}
\includegraphics{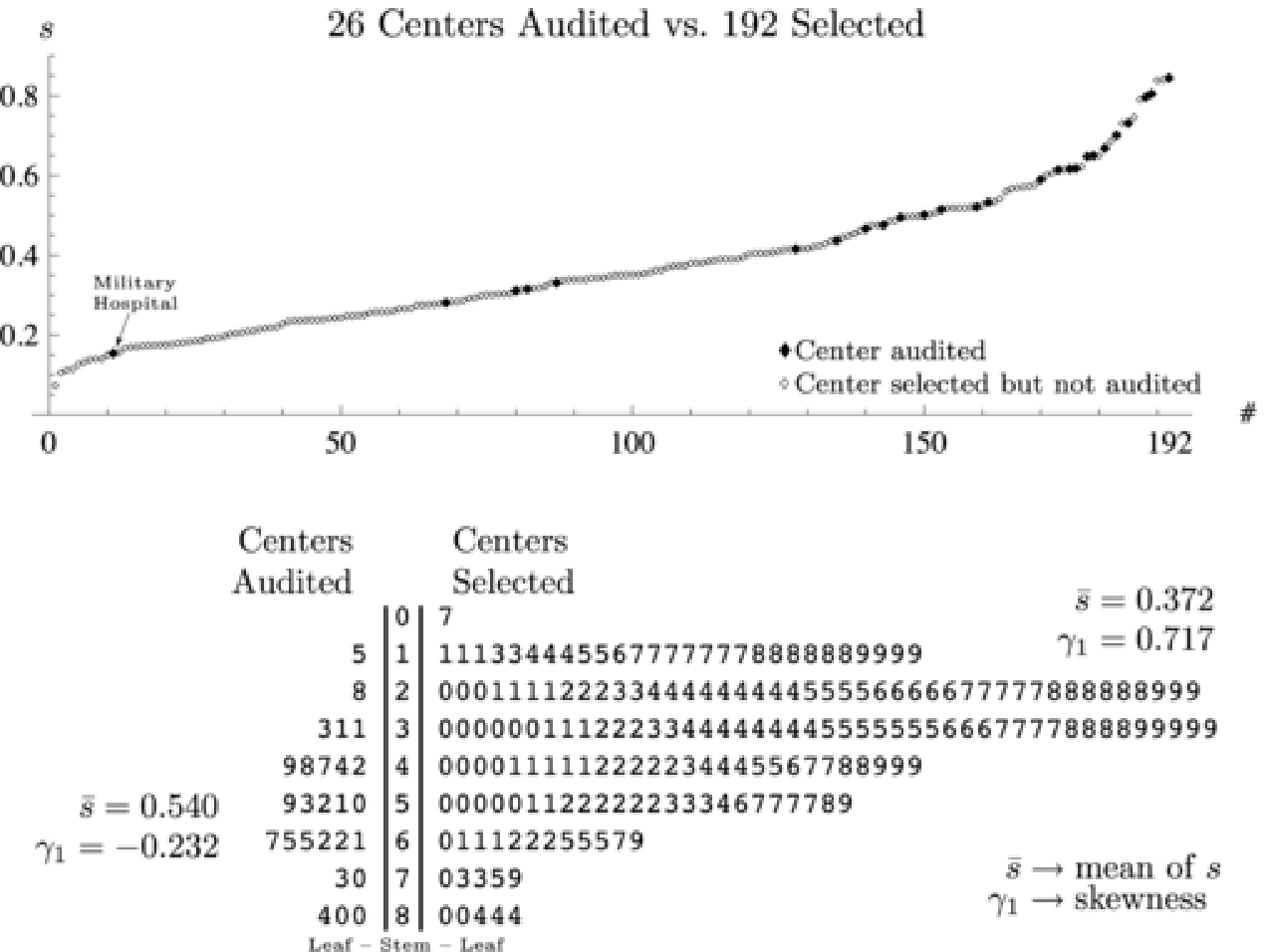}\vspace*{-1pt}
%[scale=1]{stemAndLeaf.pdf}
\caption{Comparison between the $s$ value of the 192 selected centers
and the
26 audited centers. TOP: The selected centers are ordered according to the
value of $s$ and plotted. BOTTOM: Back-to-back stem-and-leaf plot showing
the same values of $s$ as in the top figure. To obtain $s$ values, multiply
stem by 0.1 and leaves by 0.01. }\label{fig:stemAndLeaf}
\end{figure*}

Furthermore, out of 192 centers selected for hot audit, only 26 were actually
audited in the presence of witnesses representing the opposition and the
international observers. The following excerpt from the Carter Center
Comprehensive Report \cite{longCarterReport} is very illustrative:\vadjust{\goodbreak}

\begin{quote}

Auditors, table members, and military personnel were not properly informed
that the audit would occur nor were they clear about the procedure to be
followed. The instructions themselves did not clearly call for a separate
tally of the Yes and No votes, and in some centers, the auditors only counted
the total number of voters. (\ldots)

Nevertheless, Carter Center observers were able to witness six auditing
processes. In only one of the six auditing sites observed by The Carter Center
did the paper ballot receipt counting actually occur. In this place, the
auditing was conducted by the \textit{mesa} president, and the
recount of the ballots
produced exactly the same result as the \textit{acta} printed by the
voting machine. In
the rest of the sites observed, the auditor appointed by the CNE did
not allow
the opening of the ballot box, explaining his/her instructions did not include
the counting of the Yes and No ballots from multiple machines.

There were also complaints of military denying access to voting centers where
audits were being conducted. Carter Center observers could not confirm this
claim. (\ldots)

The CNE provided The Carter Center with copies of the audit reports of 25
centers. It was clear from the forms that the audit was not carried out in
many places because the fields in the form were left empty, there were no
signatures of pro-government or opposition witnesses, etc. The forms were
poorly filled out, clearly showing inadequate training. The instructions
issued by the CNE to the auditors were either incomplete or unclear.
This is a
direct consequence of issuing the audit regulation three\vadjust{\goodbreak} days before the
election. The final result was that the CNE squandered a crucial opportunity
to build confidence and trust in the electoral system and outcome of the
recall referendum.
\vspace*{1pt}
\end{quote}

Auditing only 26 centers out of 192 selected centers, is basically a
cancellation of the auditing process. But, is there anything special about
these 26 centers? If this drastic reduction in audit size was because
it was
``poorly executed,'' and poor execution is independent of the value of $s$,
then the value of $s$ of these 26 centers would behave as a random sample
within the $s$ value of the 192 selected centers.\looseness=1

From Figure~\ref{fig:stemAndLeaf}, it is clear that the 26 centers
that were
actually audited seem to have a much higher value of $s$ than the 192 centers
from where they come from. The average $s$ for the 192 selected centers is
\mbox{$\bar{s}_{\mbox{\tiny selected}}=0.372$} while for the audited ones
it is
\mbox{$\bar{s}_{\mbox{\tiny audited}}=0.540$}. Additionally, the
distribution of the
192 selected centers is positively skewed while the distribution of the 26
audited centers is negatively skewed.

Can this be just a coincidence? A Monte Carlo~simulation was done, selecting
26 random centers out of the 192 selected for auditing. The result of this
simulation is that the probability of having a~$\bar{s}_{\mbox{\tiny
audited}}=0.540$ is 1~in 50 million; and this does not take into account
the difference in skewness.

%should yield an even smaller probability.}.

This result is consistent with the hypothesis, because centers with a small
value of $s$ are the ones most susceptible to distortions.

Thus, it has been shown that the audited centers are not representative
of neither
the universe of all computerized centers, nor the restricted universe
used to
select them.

%f14 ###
\begin{figure*}%[htbp]

\includegraphics{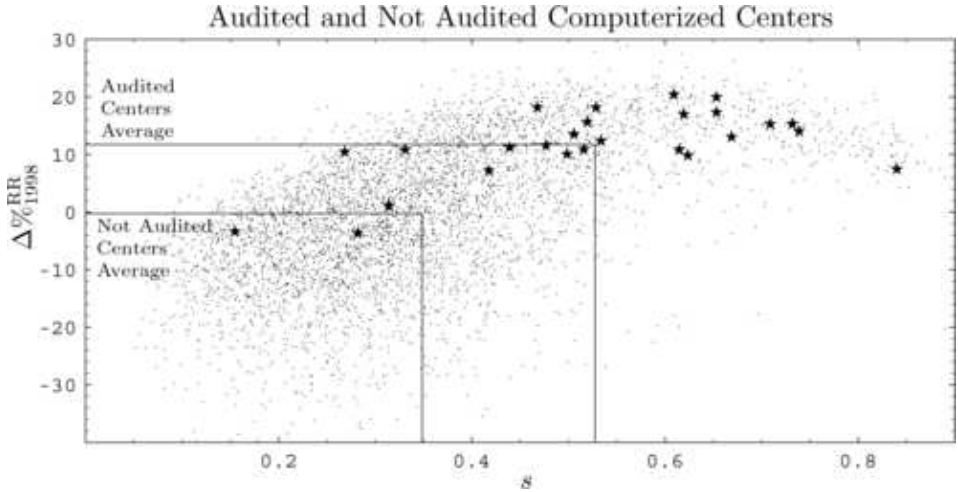}
%[scale=1]{cachito.pdf}
\caption{In this figure, the 26 computerized audited centers are
compared with
the universe of computerized centers. The average $s$ and \ensuremath
{\Delta\%_{1998}^{\mbox{\textup{\scriptsize RR}}}}\ are
indicated with lines.}\label{fig:cachito}
\end{figure*}

The audited centers are not representative of the universe of computerized
voting centers (see Figure~\ref{fig:cachito}) because:

\begin{enumerate}
\item In the audited centers, the s\'{\i} vote won by 63.47\% to
40.91\%.\vadjust{\goodbreak}
\item\ensuremath{\Delta\%_{1998}^{\mbox{\scriptsize RR}}}\ is very
different.
\item The value of $s$ is much larger.
\end{enumerate}

\begin{figure*}[t]

\includegraphics{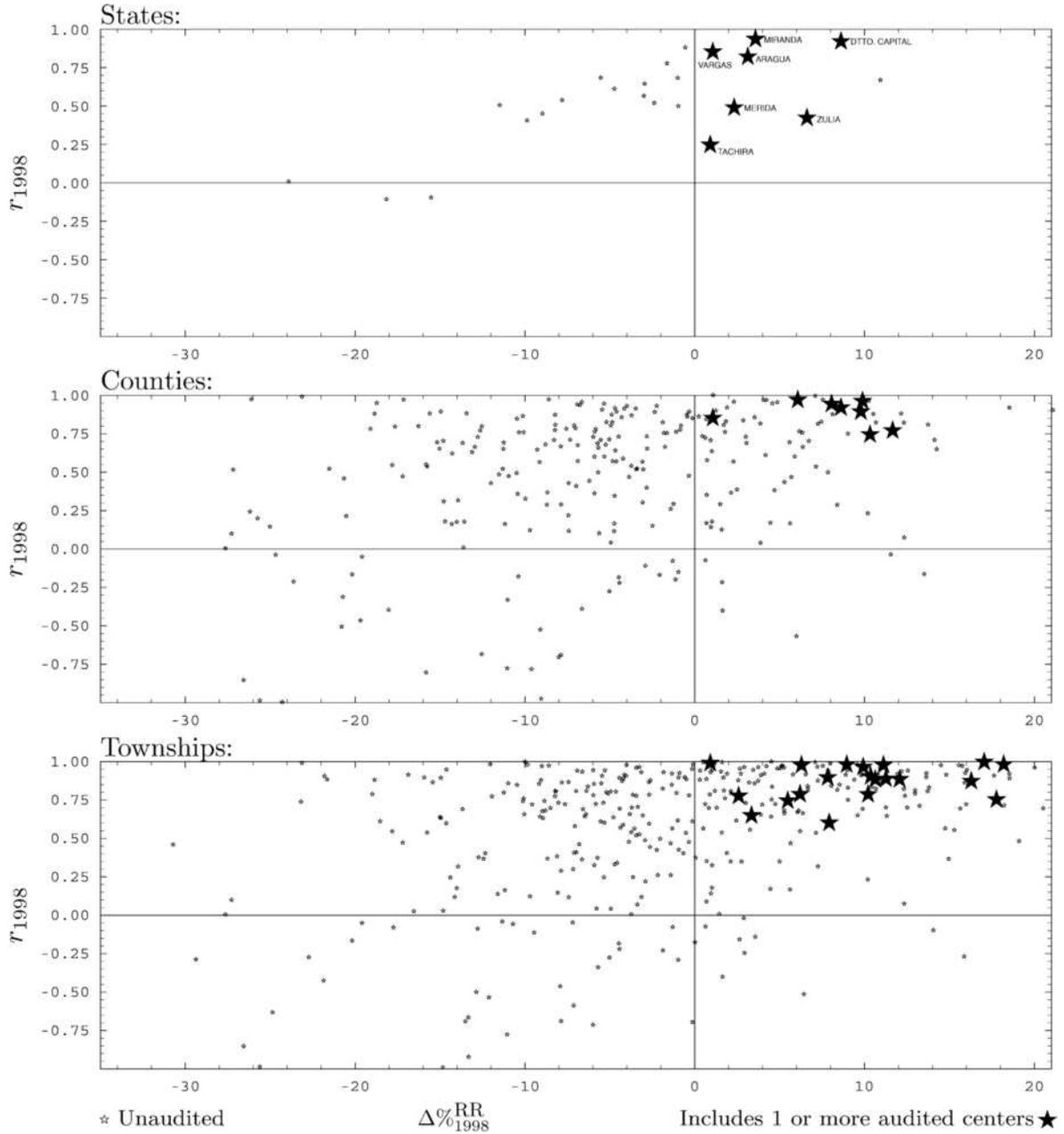}
%[scale=1]{gHotAuditTriple.pdf}
\caption{Townships, counties and states where the 26 audited centers are
located. The vertical axis is the correlation with respect to the percentage
of opposition in 1998 ($r_{1998}$). The horizontal axis is the
difference in
percentage of opposition with respect to 1998
(\ensuremath{\Delta\%_{1998}^{\mbox{\textup{\scriptsize RR}}}}).}\label
{fig:hotAuditTriple}\vspace*{3pt}
\end{figure*}

Additionally, the townships, counties and states where centers were audited
are not representative of the other townships, counties and states.
They are
not representative with regard to their \ensuremath{\Delta\%
_{1998}^{\mbox{\scriptsize RR}}}\ and the correlation with
respect to the 1998 election $r_{1998}$. This can be seen in
Figure~\ref{fig:hotAuditTriple}.

%f15 ###

%s10 ###
\section{Cold Audit}\label{sec:coldAudit}

Given the fact that the hot audit failed to serve its purpose, another audit
was made three (3) days after the referendum. This audit cannot
validate the
official results mainly because of two reasons:

\begin{itemize}

\item The audited entity itself cannot select the centers to be audited.
According to the OAS/Carter report \cite{coldaudit} ``The sample was generated
by CNE staff'' on its own computer using its own software.

\item The control mechanisms that had been implemen\-ted to certify that the
samples were unaltered were not used.

\end{itemize}

The draw to select the centers to be audited was broadcast live on
the official television station, but the results were not shown.
Usually, the whole
idea of transmitting a draw on TV, is to let the public know the
results as
they are being generated.

When the ballot boxes were closed and sealed, and the electoral centers
closed, the seal was signed by witnesses. The boxes were then taken into
custody of the military.

The following excerpt from the OAS/Carter Center report
\cite{coldaudit}\vadjust{\goodbreak}
explains the mechanism used to certify that the boxes were unaltered:

\begin{quote}
Each box was physically checked to see whether:

\begin{enumerate}
\item The material used to seal the box was intact or whether there
were signs
that it had been taken off and then replaced.
\item There were cracks or holes through\break which votes might have been extracted
or inserted.
\end{enumerate}

If a box was defective in regard to sealing, cracks, or holes, all the boxes
of that polling station were excluded from the audit and a note to that effect
recorded in the minutes.
\end{quote}

However, the witnesses who had signed the boxes were not called to
certify the
authenticity of the box.

When this audit was questioned, the Carter Center and OAS response was that:

\begin{quote}
Furthermore, the correlation between the signers and the s\'{\i} votes is almost
identical in the universe and in the sample. The difference between the
correlations is less than 1 percent:

\begin{tabular*}{195pt}{@{\extracolsep{\fill}}lc@{}}
\hline
& \textbf{Correlation coefficient}\\
\hline
Universe & 0.988\\
Sample & 0.989\\
\hline
\end{tabular*}
\end{quote}

This certainly can be used to argue that the boxes opened were representative
of the official results, but does not indicate anything in regard to
validating the official results.

Interestingly, the draws for the hot and cold audit included 16
common centers. These 16 centers were successfully cold-audited, but
none of them were allowed to be hot-audited.

%s11 ###
\section{Conclusions}

We have explored the voting data arising from the RR carried out in
2004 and
also the results of two audits conducted after the RR took place. We have
identified several issues associated with the results obtained from
voting centers
using touch-screen voting machines. In particular:

\begin{enumerate}
\item The official s\'{\i} results in computerized centers seem to behave in
an excessively linear fashion relative
to the number of signatures in support of the RR in each voting center (see
Section~\ref{sec:signatures}).
\item The official s\'{\i} results in computerized centers are surprising
given the results
of the 1998 elections in those same centers (see Section~\ref{sec:1998}).
\item The percentage of votes for the opposition seem to be too highly
correlated
with $s$, the relative~num\-ber of signatures in a voting center, in
particular in
those centers where $s$ was small (see Section~\ref{sec:rDiff}).\vspace*{-2pt}
\end{enumerate}

When combined with the facts that in general, paper ballots were not counted
and that voting machines were connected to a central CNE server \emph
{before} voting
results could be printed, these observations suggest that the official results
obtained from computerized voting centers deserve a closer look.

In principle, two audits---a hot audit carried out immediately
following the
referendum and a cold audit carried out three days later---should have helped
resolve any questions arising about the voting and vote counting processes.
However, an analysis of the data that resulted from the two audits
reveals that
the audits were not conducted as had originally been announced and thus could
not alleviate doubts about the official results nor could they be used
to certify the
accuracy of results. In particular, we argue that:\vspace*{-2pt}

\begin{enumerate}
\item The computerized centers in the 20 counties to which the hot
audit was
restricted by the CNE were not representative of the universe of
computerized voting
centers (Figure~\ref{fig:20counties}).
\item The hot-audited centers were not representative of the rest of the
computerized centers (Figure~\ref{fig:cachito}).
\item Townships, counties and states where computerized centers were hot-audited were not a representative sample of townships, counties and
states in
Venezuela (Figure~\ref{fig:hotAuditTriple}).
\item The probability that the centers that were hot-au\-dited do not
appear to be a
random sample~of all computerized voting centers seems to be high and thus
it is difficult to believe that the unexpect\-ed sample of audited
centers was due
to chance alone. Note that centers that were actually audited were drawn
from a subsample of all centers with a high proportion of signatures
(Figure~\ref{fig:stemAndLeaf}).\looseness=-1
\item Audits were suspended in centers with low $s$, where the
linearity in the official
results is most questionable.\vspace*{-2pt}
\end{enumerate}

While none of this constitutes proof of tampering, we believe that our
analyses of
some of the data collected in association with the recall referendum
cast some
doubt about the accuracy of the official results. If in fact it is
reasonable to
assume that:\vspace*{-1pt}

\begin{itemize}
\item A person who signed the form requesting a referendum was likely
to vote s\'{\i}.\vadjust{\goodbreak}
\item A person who did not sign the form is not necessarily likely to
vote no,
then the very high correlation between the proportion of signers and the
proportion of s\'{\i} votes at a center should be viewed with suspicion
rather than as
a confirmation that official results are believable, as the OAS/Carter Center
report claim. Indeed, an excerpt from the report states that:\vspace*{-2pt}%
\begin{quote}
``There is a \textit{high correlation between the number of YES votes per
voting center and the number of signers of the presidential recall
request per
voting center}; the places where more signatures were collected also
are the
places where more YES votes were cast. There is no anoma\-ly in the
characteristics of the YES votes when compared to the presumed
intention of
the signers to recall the \mbox{president}.''\looseness=-1\vspace*{-2pt}%
\end{quote}
\end{itemize}

We argue exactly the opposite and have provided persuasive arguments to
support our position.

\begin{appendix}

%s12 ###
\vspace*{-3pt}\section{Data Processing Methodology}\vspace*{-3pt}

Official Referendum results were downloaded\break from~the CNE website:
\texttt{\href{http://www.cne.gob.ve/referendum\_presidencial2004/}{http://www.cne.gob.ve/}\break
\href{http://www.cne.gob.ve/referendum\_presidencial2004/}{referendum\_presidencial2004/}}.

The download was automated using a \mbox{custom-made} Perl script.
All the data was stored on a MySQL~data\-base. Calculations were made
using \textit{Mathematica 5.2} which was connected to MySQL using the Database\-Link package.
Electoral results from the 1998 presidential election
were obtained on an original CNE CD-ROM, and the data was converted
from Microsoft
Access to MySQL. The REP from July 2004 was also converted from MS
Access to
MySQL. The CNE signature data was obtained on a CD from S\'umate, and
is the
same version given to the OAS and the Carter Center. This data was
supplied in a~single text file.

\renewcommand{\thefigure}{\arabic{figure}}
\setcounter{figure}{15}
%f16 ###
\begin{figure*}[t]

\includegraphics{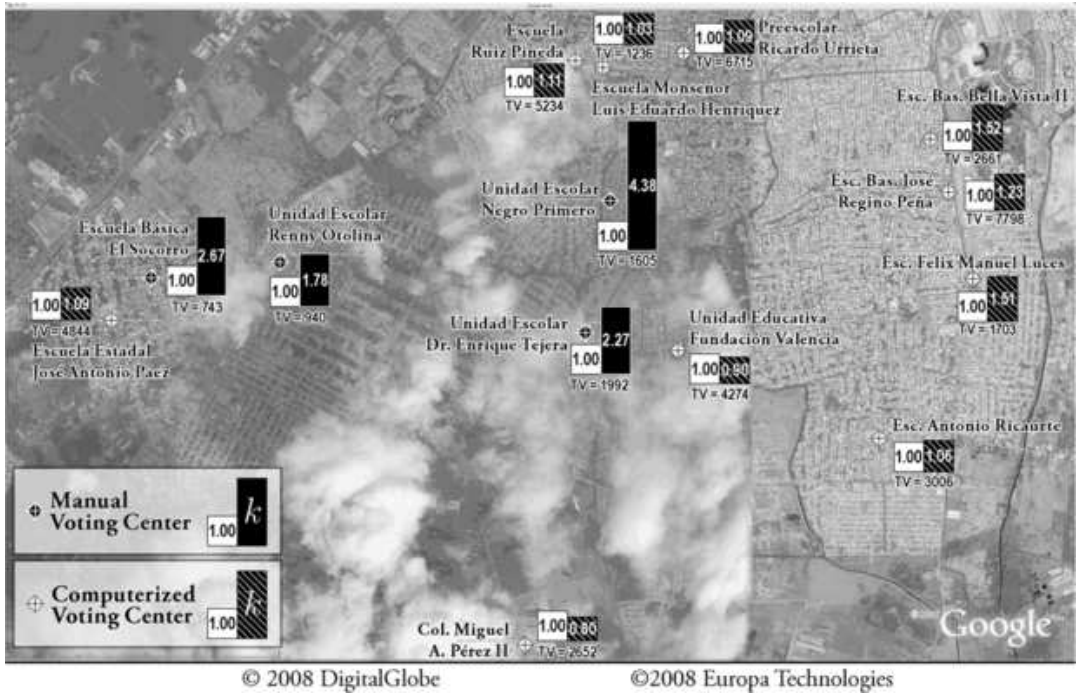}\vspace*{-4pt}
%[]{google-bw.pdf}
\caption{Partial aerial view of Miguel Pe\~na township (taken from the
\emph{Google Earth\texttrademark} mapping service). Manual versus
computerized voting centers are compared in regards to their $k$ value
and total number of votes (TV). The image is centered at Latitude
$10^\circ7'32.66''$N and Longitude $68^\circ1'22.48''$W. }\label{fig:google}
\vspace*{-4pt}
\end{figure*}

By matching people's identification numbers (c\'e\-dula number) from the
signatures and REP
data, it was possible to find the number of signatures per voting center.

In order to classify voting centers into manual and computerized, the
following sources of information were used:\vspace*{-3pt}

\begin{itemize}
\item S\'umate's list of computerized and manual voting
centers.
\item On the CNE website, computerized centers show results down to
the voting
machine level, whereas manual voting centers show results down to the
voting table
level.
\end{itemize}

The list of computerized and manual centers obtained using the aforementioned
sources was compared on a township by township basis with the CNE
infrastructure document \cite{infraestructura}.

The list of centers effectively audited on the day of the Referendum was
obtained from a document given by the \textit{Coordinadora Democr\'
atica} to the
Carter Center and OAS. A copy of this document and the data needed to
reproduce this study can be found at:
\url{http://esdata.info/2004}.

\renewcommand{\thefigure}{\arabic{figure}}
\setcounter{figure}{16}
%f17 ###
\begin{figure*}[t]

\includegraphics{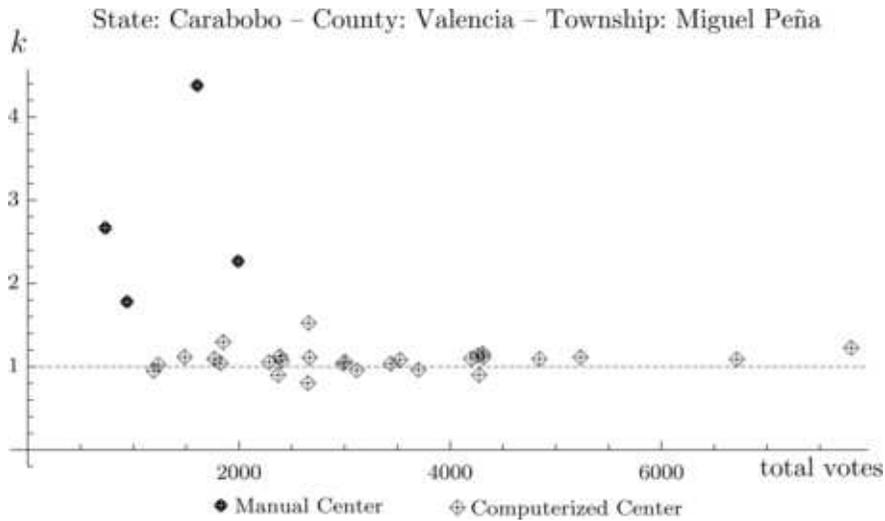}
%[]{k_vs_total_votes.pdf}
\caption{Behavior of $k$ versus total votes in all of Miguel Pe\~na's
voting centers. Manual and computerized centers are
shown.}\label{fig:google2}\vspace*{-3pt}
\end{figure*}

\renewcommand{\thefigure}{\arabic{figure}}
\setcounter{figure}{17}
%f18 ###
\begin{figure*}[b]
\vspace*{-3pt}
\includegraphics{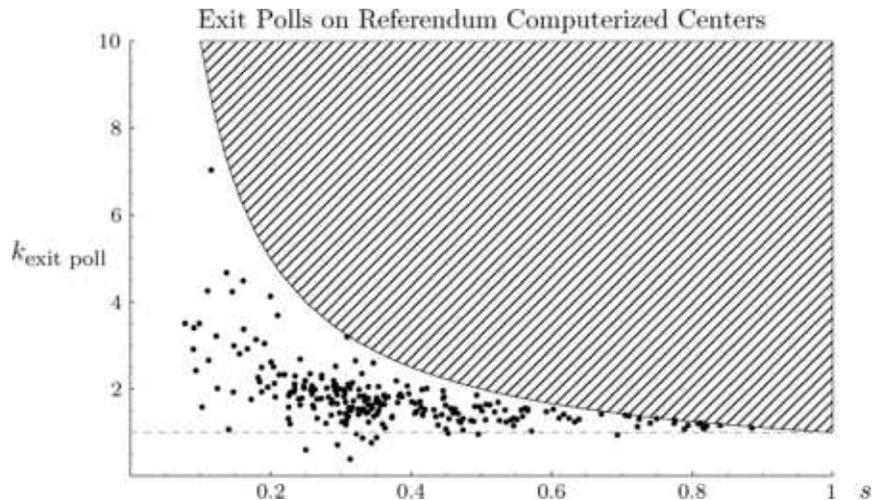}
%[]{laurentEP.pdf}
\caption{Exit polls at computerized centers.}\label{fig:laurentEP}
\end{figure*}

%f19 ###
\begin{figure*}
\includegraphics{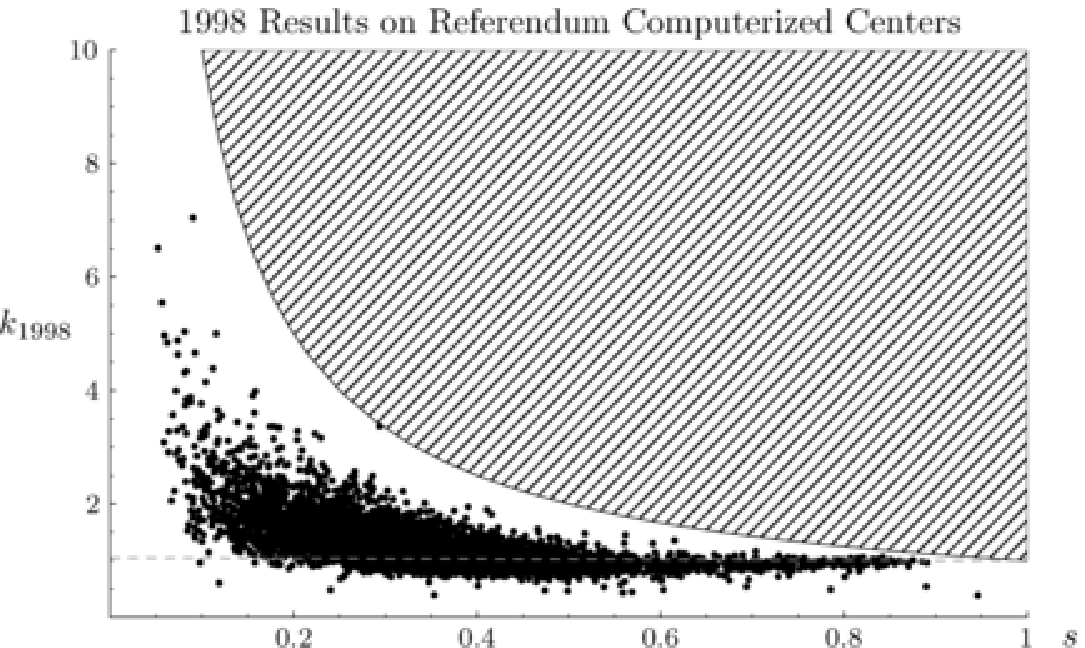}
%[]{laurent1998.pdf}
\caption{1998 presidential election at computerized centers.}\label
{fig:laurent1998}
\end{figure*}

The coordinates of the voting centers shown in Appendix~\ref
{apndx:google} were provided by ``Delta Electoral.''

The simulation was done using a deck of cards shuffling algorithm. The random
number generator used by this algorithm was the ``Wolfram rule 30 cellular
automaton generator for integers,'' which is provided by \textit{Mathematica}.

%s13 ###
\section{A Mixed Township Example}\label{apndx:google}

Miguel Pe\~na is a township in Valencia County, in the state of
Carabobo. It is one of the townships with higher population in the
country. It had 32 voting centers, 28 computerized and 4
manual.

In Figure~\ref{fig:google}, a partial aerial view of this township is
shown. In it, notice that manual and computerized voting centers are in
the same urban neighborhood. Despite this, the values of $k$ are much
higher for the manual centers than for the surrounding computerized
centers, regardless of the total number of votes.

In Figure~\ref{fig:google2} notice that in this township, the lowest
$k$ value of the 4 manual centers is greater than the maximum $k$ value
of the 28 computerized voting centers.

\renewcommand{\thefigure}{\arabic{figure}}
\setcounter{figure}{19}
%f20 ###
\begin{figure*}[b]
\includegraphics{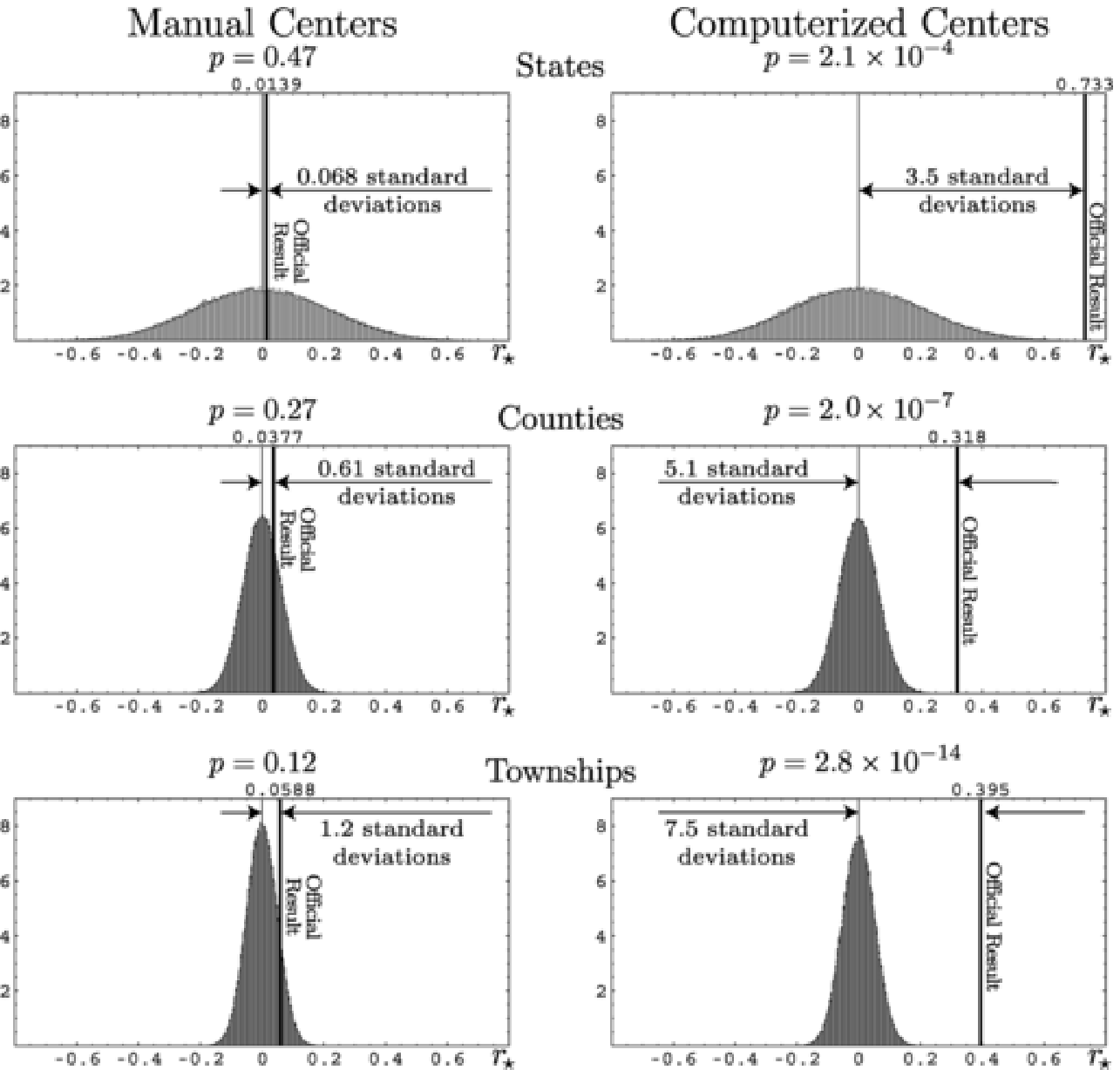}
%[]{sixMonteCarlo.pdf}
\caption{
Comparison of official results correlation $r_{\!\star}$ versus
expected value
distribution found after 100,000 simulations for manual and computerized
centers at state, county or township level. The simulation results
follow a
normal distribution, which is shown as a dotted line. The probability
of the
official $r_{\!\star}$ happening by chance is indicated as
$p$.}\label{fig:sixMonteCarlo}
\end{figure*}

%s14 ###
\section{Additional Nonlinearity Plots}

According to the exit polls made under the supervision of Penn, Schoen and
Berland Associates, the opposition won the Referendum by a wide margin. By
changing the numerator of equation~(\ref{eq:singularidad}) from
percentage of s\'{\i}
votes to percentage of s\'{\i} from exit polls, a~value of $k_{\mbox
{\footnotesize
exit polls}}$ can be calculated. The result, for computerized centers
only, is
plotted on Figure~\ref{fig:laurentEP}.

Similarly, $k_{1998}$ can be calculated by using the percentage of opposition
in the 1998 presidential election in the numerator of
equation~(\ref{eq:singularidad}). The result, for computerized centers
only, is
shown on Figure~\ref{fig:laurent1998}.

%s15 ###
\section{\texorpdfstring{Monte Carlo Simulations for Correlation Between \ensuremath
{\Delta\%_{1998}^{\mbox{\scriptsize RR}}}\ and
\lowercase{$r_{1998}$}.}
{Monte Carlo Simulations for Correlation Between Delta\%_{1998}^{RR} and
r_{1998}}}\label{apndx:mcarlo}

Assuming that \ensuremath{\Delta\%_{1998}^{\mbox{\scriptsize RR}}}\
and $r_{1998}$ are independent, regardless of being
calculated at state, county or township level, then the correlation between
them~$r_{\!\star}$ must be casual. In order to find the probability
that the
observed $r_{\!\star}$ is casual, it is possible to reorder the values of
$r_{1998}$ with respect to \ensuremath{\Delta\%_{1998}^{\mbox
{\scriptsize RR}}}. This reordering was made 100,000
times and the correlation was calculated each time. In all cases, the
resulting distribution was found to be normal. The estimated
probabilities for
manual and computerized centers at state, county or township level are shown
in Figure~\ref{fig:sixMonteCarlo}.

%s16 ###
\section{Differences in Characteristics, Official Results and REP
Variation of
the 20 Counties Subject to Hot-Audit Drawing in Comparison to the Other
Counties}\label{apndx:k20counties}

When the CNE decided to restrict the audit to 20 urban counties, it created
two groups of computerized centers:

\begin{itemize}
\item2,040 computerized centers inside the 20 counties and therefore
subject to be selected in the draw. Variables referring to these
centers will use a 20 as a subindex ($\bullet_{20}$).
\item2,553 computerized centers not subject to hot audit at
all. Variables referring to these centers will use a
$\varnothing$ as a subindex ($\bullet_{\varnothing}$).\vadjust{\goodbreak}
\end{itemize}

\renewcommand{\thefigure}{\arabic{figure}}
\setcounter{figure}{20}
%f21 ###
\begin{figure*}[t]

\includegraphics{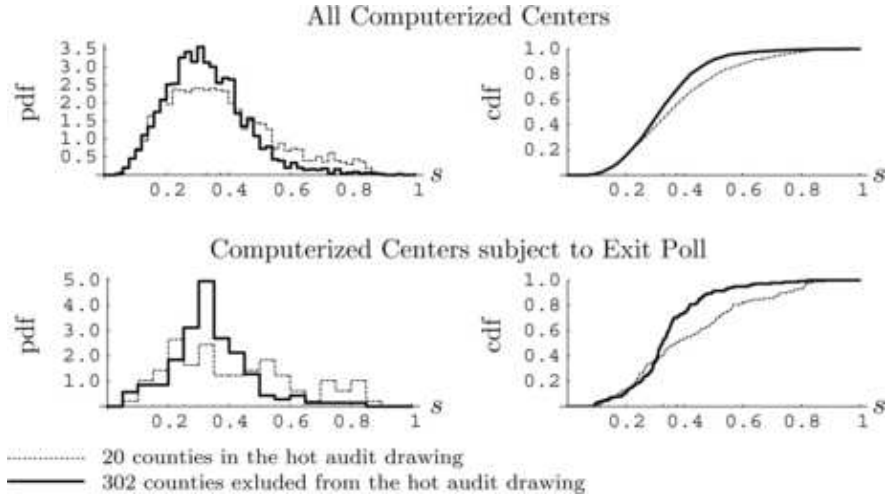}
%[]{sDist.pdf}
\caption{Comparison of $s$ probability density function (pdf) and cumulative
density function (cmf) for computerized centers inside the 20 counties
of the
hot audit and in the 302 excluded counties.}\label{fig:sDist}
\vspace*{3pt}
\end{figure*}

In Figure~\ref{fig:20counties} it is shown that the behavior in computerized
centers in the 20 counties is very different from that of the rest of the
country.

%s16.1 ###
\subsection{Differences in Characteristics}

When the CNE set up the signature collection event, it established the number
of signature collection centers (SCC) directly in proportion to the
number of
people in the electoral registry (REP) for each county. A~lot of people live
in urban counties, therefore, a~lot of SCCs were assigned to these counties.
Thus, access from where the people lived to where they had to sign was much
easier in these 20 counties. On the other hand, voting centers are more
numerous and better distributed throughout the national territory.

For example, a county like Chacao in the Miranda state has 27 $\mbox
{km}^2$ of
area and 11 SCCs. In Chacao there were 24 voting centers, all of them
computerized. On the other hand, the much larger Macanao Pen\'\i nsula in
Margarita Island has an area of 330.7 $\mbox{km}^2$ and only had 3
SCCs. There
were 8 voting centers in Macanao, all of them computerized.\looseness=1

\renewcommand{\thefigure}{\arabic{figure}}
\setcounter{figure}{21}
%f22 ###
\begin{figure*}
\includegraphics{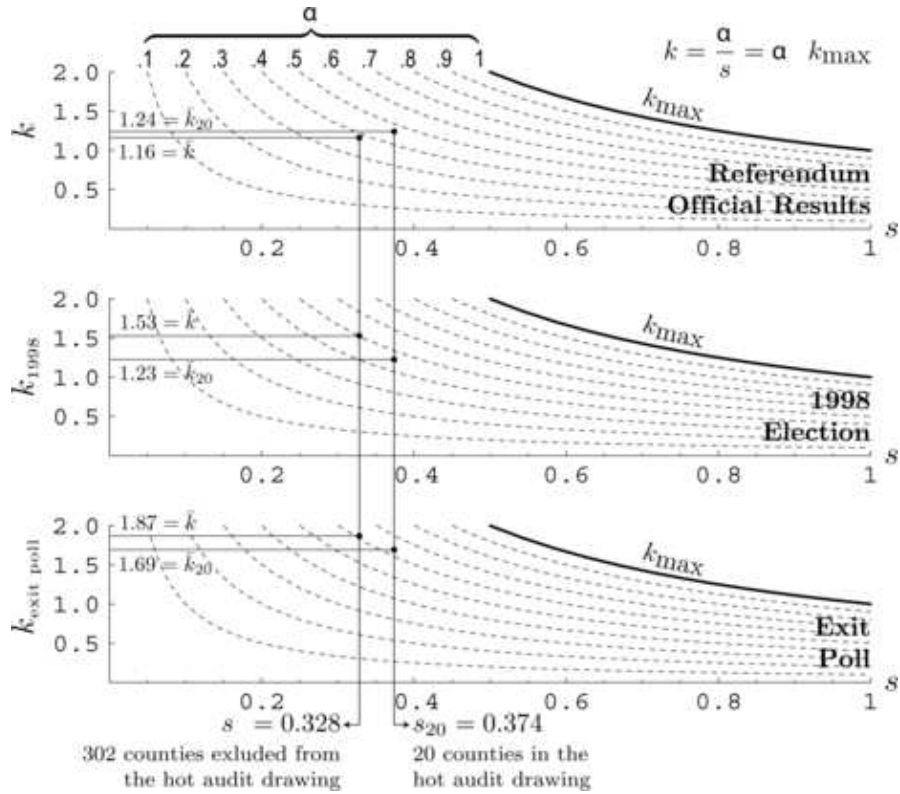}
%[]{two_points.pdf}
\caption{
Comparison of average $k$ and $s$ values for the computerized centers inside
and outside the 20 counties to which the hot-audit universe was restricted.
These $\bar{k}$ and $\bar{s}$ values are shown for the official referendum
results, for the 1998 presidential election and for the referendum exit polls.
}\label{fig:ballena}
\end{figure*}

%f23 ###
\begin{figure*}%[b]

\includegraphics{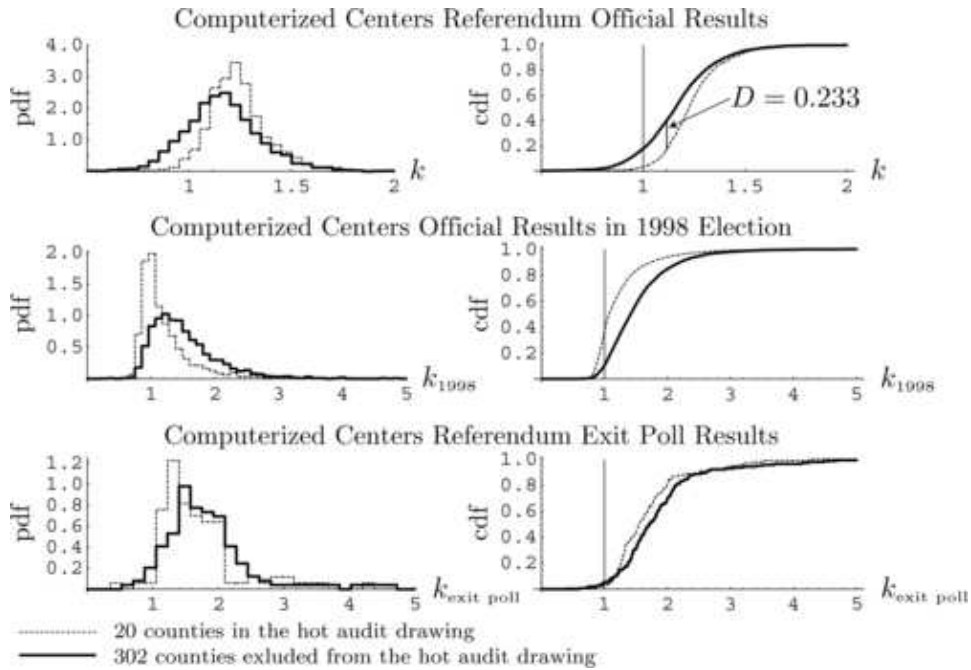}
%[]{kDist.pdf}
\caption{Comparison of $k$ probability density function (pdf) and cumulative
density function (cmf) for computerized centers inside the 20 counties
of the
hot audit and in the 302 excluded counties. The maximum cmf difference
(Supremum) for the official results is shown as $D$.}\label{fig:kDist}
\end{figure*}

\renewcommand{\thefigure}{\arabic{figure}}
\setcounter{figure}{23}
%f24 ###
\begin{figure*}[t]

\includegraphics{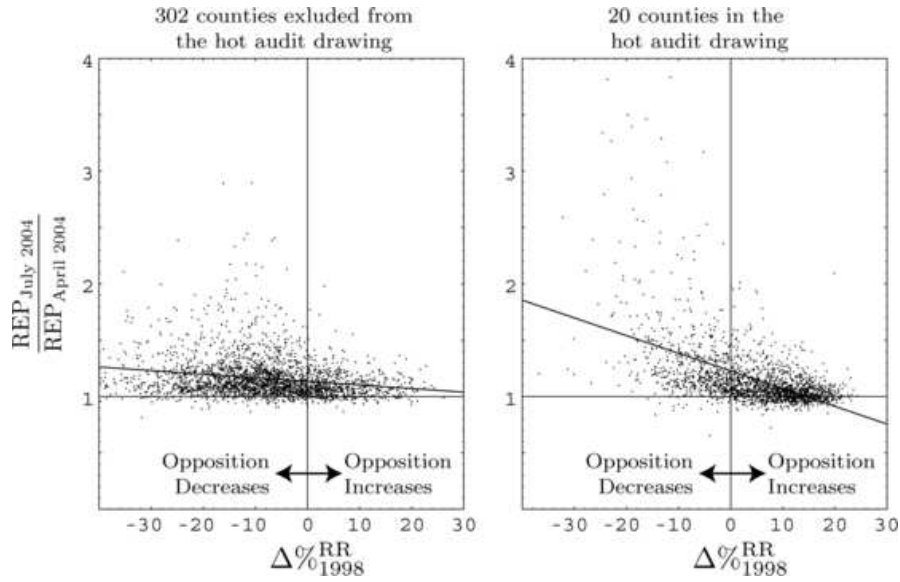}
%[]{rep2.pdf}
\caption{REP variation versus \ensuremath{\Delta\%_{1998}^{\mbox
{\textup{\scriptsize RR}}}}\ in computerized centers inside and
outside the 20 counties of the hot-audit drawing. A least-square line is
included in both cases. }\label{fig:rep2}\vspace*{-4pt}
\end{figure*}
%
%f25 ###
\begin{figure*}[b]
\vspace*{-4pt}
\includegraphics{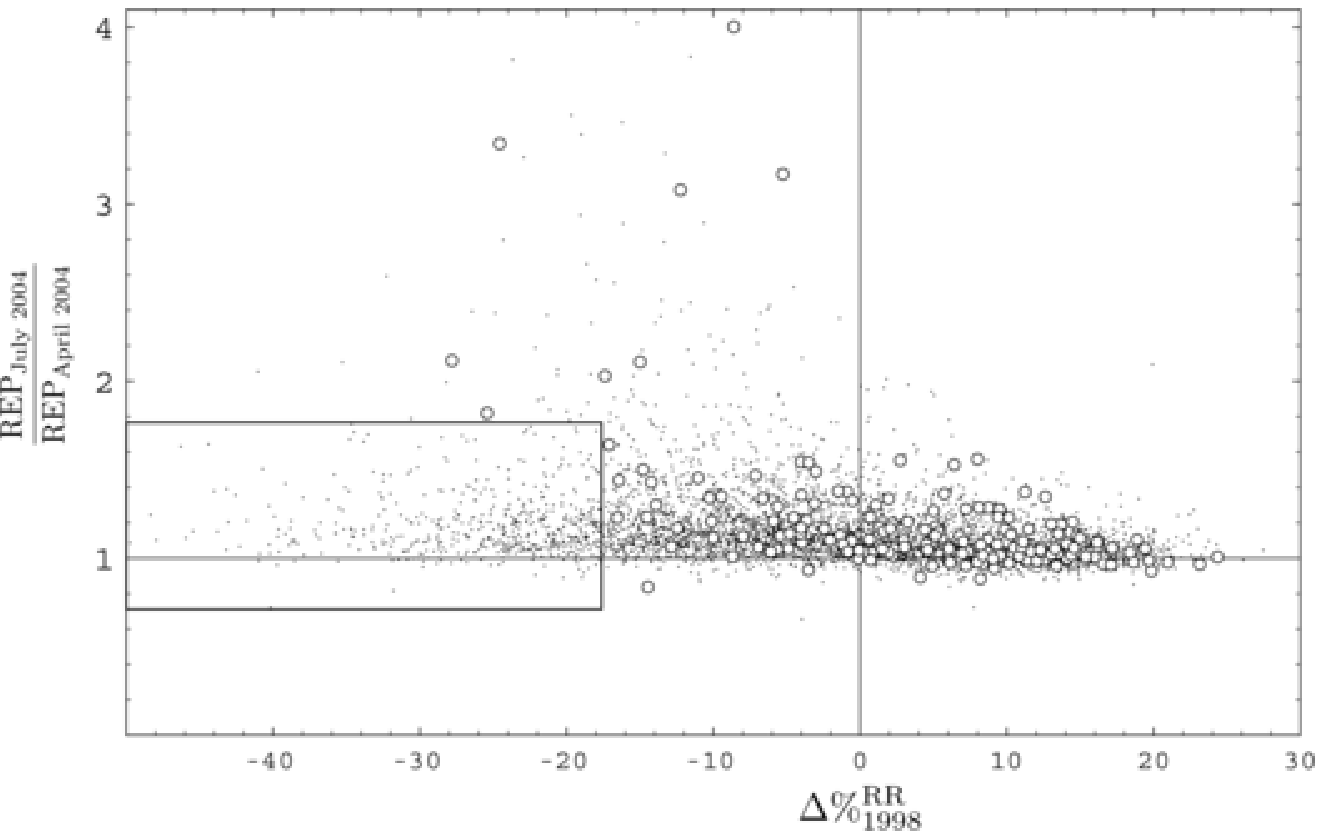}
%[]{rep3.pdf}
\caption{REP variation versus \ensuremath{\Delta\%_{1998}^{\mbox
{\textup{\scriptsize RR}}}}\ in all computerized centers indicating
the 192 selected for hot-auditing. None of the 192 selected centers
were in
the rectangle area.}\label{fig:rep3}
\end{figure*}

In Figure~\ref{fig:sDist}, it can clearly be seen that the 20 counties
have higher $s$ values which is consistent with the ideas just explained.

%This can be seen in Figure~\ref{fig:sDist}, it is clear that the 20
%counties
%have higher $s$ values, which is consistent with the ideas just
%explained.

There were many computerized centers in rural areas where it was much more
difficult to sign than to vote. When the audit universe was restricted
to 20
\textit{urban} counties, all computerized centers in rural areas, the
ones with
a higher uncertainty in $k$, were excluded from the hot-audit drawing
universe.\vadjust{\goodbreak}

%s16.2 ###
\subsection{Differences in Results}

When the value of $s$ decreases, in general, it is expected that the $k$
values should increase, after all, $k_{\scriptsize{max}}=1/s$. Hence, a
larger $k$
is expected in rural counties than in the 20 counties of the hot
audit where signing was less troublesome. However, in the official results,
exactly the opposite occurred, as shown in Figure~\ref{fig:ballena}.

Considering that for the official referendum results~$\bar{k}_{20}$ is the
average of 2,040 voting centers and~$\bar{k}_{\varnothing}$ is the
average of the
remaining 2,553 voting centers, how likely is it that just by chance,
$\bar{k}_{20}$ be larger than~$\bar{k}_{\varnothing}$ by 3.4\%? What
could be
expected is that~$\bar{k}_{20}$ would be smaller than $\bar
{k}_{\varnothing}$.
Contrary to official results, in the exit polls and in the 1998 election
$\bar{k}_{20}$ is significantly less than $\bar{k}_{\varnothing}$, as
shown in
Figure~\ref{fig:ballena}.

As seen in Figure~\ref{fig:kDist}, the distribution of $k$ values
among the
2,040 auditable centers is quite different from that of the 2,553
nonauditable
centers. The~$k$ values in the 2,040 auditable centers tend to be
larger than
in the other 2,553 nonauditable centers. The portion of centers with $k$
smaller or near to 1, is much smaller in the 2,040 auditable centers
than in
the other 2,553. That is contrary to what happened in the 1998
election and in
the exit poll. Additionally, note that the $k$~pdf seems to be much more
symmetric than that in the 1998 results or the exit polls.

How likely is it that $k_{20}$ cmf be below $k_\varnothing$ cmf with
such a
large difference ($D=0.233$)? Being conservative and assuming that both
$k_{20}$ and $k_{\varnothing}$ distri\-butions came from the same continuous
distribution, the probability can be estimated using the Kolmogo\-rov--Smirnov
Test for two samples. This probability was found to be in the order of
$2.6\times10^{-54}$. For the reasons previously exposed, the
distribution of
$k_\varnothing$ should be greater---not equal---than that of~$k_{20}$. Hence,
the actual probability should be much smaller.

%s16.3 ###
\subsection{Electoral Registry (REP) Differences}\vspace*{-3pt}

Between April and July 2004, 1,842,959 (14.9\%) voters were added to the
REP.\vadjust{\goodbreak} In the computerized centers the number of registered voters went
from 10,849,321
to 12,390,159. In Figure~\ref{fig:rep2} it is shown how differently these
increments were distributed in the computerized centers. Furthermore, in
Figure~\ref{fig:rep3}, it can be seen that the 192 centers selected to
be hot
audited exclude an area where the government has important gains
without a big
increase in the REP.
\end{appendix}

\section*{Acknowledgment}

The authors would like to thank all the people who contributed and supported
this effort. Countless hours of volunteer work helped make this paper a
reality.%\newpage

\end{document}